\documentstyle[prl,epsf,floats,aps,amsfonts,twocolumn,eqsecnum]{revtex}

\begin{document}

\twocolumn[\hsize\textwidth\columnwidth\hsize\csname
@twocolumnfalse\endcsname

\title{Exact spectra of conformal supersymmetric nonlinear sigma models\\
in two dimensions}
\author{N. Read$^1$ and H. Saleur$^{2,3}$}
\address{$^1$ Department of Physics, Yale
University, P.O. Box 208120, New Haven, CT 06520-8120\\ $^2$
Department of Physics, University of Southern California, Los
Angeles, CA 90089-0484\\ $^3$ CIT-USC Center for Theoretical
Physics, USC, Los Angeles CA 90089-2535}
\date{June 14, 2001}
\maketitle

\begin{abstract}
We study two-dimensional nonlinear sigma models in which the
target spaces are the coset supermanifolds
{U($n+m|n$)/[U($1)\times$U($n+m-1|n$)]} {$\cong{\bf
CP}^{n+m-1|n}$} (projective superspaces) and
{OSp($2n+m|2n$)/OSp($2n+m-1|2n$)} {$\cong S^{2n+m-1|2n}$}
(superspheres), $n$, $m$ integers, $-2\leq m\leq 2$; these quantum
field theories live in Hilbert spaces with indefinite inner
products. These theories possess non-trivial conformally-invariant
renormalization-group fixed points, or in some cases, lines of
fixed points. Some of the conformal fixed-point theories can also
be obtained within Landau-Ginzburg theories. We obtain the
complete spectra (with multiplicities) of exact conformal weights
of states (or corresponding local operators) in the isolated
fixed-point conformal field theories, and at one special point on
each of the lines of fixed points. Although the conformal weights
are rational, the conformal field theories are not, and (with one
exception) do not contain the affine versions of their
superalgebras in their chiral algebras. The method involves
lattice models that represent the strong-coupling region, which
can be mapped to loop models, and then to a Coulomb gas with
modified boundary conditions. The results apply to percolation,
dilute and dense polymers, and other statistical mechanics models,
and also to the spin quantum Hall transition in noninteracting
fermions with quenched disorder.

\end{abstract}

\pacs{PACS numbers: } ]

%%%%%%%%%%%%%%%%%%%%%%%%%%%%%%%%%%%%%%%%%%%%%%%%%%%%%%%%%%%
\section{Introduction}
\label{intro}

In this paper we consider two-dimensional (2D) nonlinear sigma
models, and also other related quantum field theories, that have
{\bf Z}$_2$-graded Lie algebras (superalgebras) as internal
symmetries. Such models have arisen in various contexts: the
integer quantum Hall (IQH) effect
\cite{pruiskenrev,efetov,pruisken1,wz,cc,read,z,km}, and recent
generalizations
\cite{az,sfbn,kag,sf1,glr,cardy,bcsz,sf,readgr,bsz,grl,rl,crkhal,mc},
as well as other problems of fermions with quenched disorder
\cite{gweg,gade,hs,altsi,zirn99,gll}; percolation, polymers, and
other statistical mechanics problems (Refs.\ \cite{parsou,glr} and
this paper); and strings in anti-de Sitter space (reviewed in
Ref.\ \cite{maldrev}, and see Refs.\ \cite{berk,bersh}). In the
models on which we concentrate here, the target spaces are
particular Riemannian supersymmetric spaces \cite{az}, in cases in
which the underlying manifolds are compact. These spaces are
U($n+m|n$)/[U($1)\times$U($n+m-1|n$)] $\cong{\bf CP}^{n+m-1|n}$
(projective superspaces) and OSp($2n+m|2n$)/OSp($2n+m-1|2n$)
$\cong S^{2n+m-1|2n}$ (superspheres). In most cases (in a suitable
range for $m$, and for $n$ sufficiently large), the beta function
for the coupling in the nonlinear sigma model is nonzero, and
there is a single non-trivial renormalization-group (RG)
fixed-point theory for each model. We make use of lattice models
that we argue have transitions in the same respective universality
classes to obtain the exact complete spectra of energy and
momentum eigenvalues (i.e.\ conformal weights) of states and the
corresponding local operators, with their multiplicities, in the
conformal field theories (CFTs) that describe the RG fixed points.
In two cases, the beta function in the sigma model is identically
zero nonperturbatively, and we find the exact spectrum at one
special point on each of these two lines of fixed-point theories.
In some cases, there are also supersymmetric Landau-Ginzburg
theories \cite{parsou} that have a transition in the same
universality class.

In more detail, the field theories and their critical points that
we discuss are as follows. For the case of ${\bf CP}^{n+m-1|n}$,
the fields can be represented by complex components $z^a$ ($a=1$,
\ldots, $n+m$), $\zeta^\alpha$ ($\alpha=1$, \ldots, $n$), where
$z^a$ is commuting, $\zeta^\alpha$ is anticommuting. In these
coordinates, at each point in spacetime, the solutions to the
constraint $z_a^\dagger z^a+\zeta_\alpha^\dagger \zeta^\alpha=1$
(we use the conjugation $\dagger$ that obeys
$(\eta\xi)^\dagger=\xi^\dagger\eta^\dagger$ for any $\eta$,
$\xi$), modulo U(1) phase transformations $z^a\mapsto e^{iB}z^a$,
$\zeta^\alpha\mapsto e^{iB}\zeta^\alpha$, parametrize ${\bf
CP}^{n+m-1|n}$. The Lagrangian density in two-dimensional
Euclidean spacetime is
\begin{eqnarray}
{\cal L}&=&\frac{1}{2g_\sigma^2}\left[(\partial_\mu
-ia_\mu)z_a^\dagger (\partial_\mu+ia_\mu)z^a\right.
\nonumber\\&&\left.\mbox{}+(\partial_\mu
-ia_\mu)\zeta_\alpha^\dagger
(\partial_\mu+ia_\mu)\zeta^\alpha\right]\nonumber\\
&&\mbox{}+\frac{i\theta}{2\pi}(\partial_\mu a_\nu-\partial_\nu
a_\mu),\label{cpnlag}
\end{eqnarray}
where $a_\mu$ ($\mu=1$, $2$) stands for the combination
\begin{equation}
a_\mu=\frac{i}{2}[z_a^\dagger\partial_\mu z^a+\zeta_\alpha
^\dagger\partial_\mu \zeta^\alpha-(\partial
z_a^\dagger)z^a-(\partial \zeta_\alpha^\dagger)\zeta^\alpha],
\end{equation}
and we work on a rectangle with periodic boundary conditions on
$z^a$, $\zeta^\alpha$. The fields are subject to the constraint,
and under the U(1) gauge invariance, $a_\mu$ transforms as a gauge
potential; a gauge must be fixed in any calculation. This set-up
is similar to the nonsupersymmetric ${\bf CP}^{m-1}$ model in
Refs.\ \cite{dadda,wit}. The coupling constants are $g_\sigma^2$,
the usual sigma model coupling (there is only one such coupling,
because the target supermanifold is a supersymmetric space, and
hence the metric on the target space is unique up to a constant
factor), and $\theta$, the coefficient of the topological term, so
$\theta$ is defined modulo $2\pi$.

First we note a well-known important point about the
supersymmetric models: the physics is the same for all $n$, in the
following sense. For example, in the present model, correlation
functions of operators that are local functions (possibly
including derivatives) of components $a\leq n_1+m$, $\alpha\leq
n_1$, exist for $n=n_1$, say, and are equal to those of the same
operators for any other value $n>n_1$, due to cancellation of the
``unused'' even and odd index values. This can be seen in
perturbation theory because the ``unused'' index values appear
only in summations in closed loops, and their contributions
cancel, but is also true nonperturbatively (it can be shown in the
lattice constructions we discuss below). In particular, the
renormalization group (RG) flow of the coupling $g_\sigma^2$ is
the same as for $n=0$, a non-supersymmetric sigma model. For the
case of ${\bf CP}^{n+m-1|n}$, the perturbative beta function is
the same as for ${\bf CP}^{m-1}$, namely (we will not be precise
about the normalization of $g_\sigma^2$)
\begin{equation}
\frac{dg_\sigma^2}{dl}=\beta(g_\sigma^2)=mg_\sigma^4+O(g_\sigma^6)
\label{umbet}
\end{equation}
where $l=\ln L$, the logarithm of the length scale at which the
coupling is defined [see e.g.\ Ref.\ \cite{wegner}, eq.\ (3.4)].
(The beta function for $\theta$ is zero in perturbation theory,
and that for $g_\sigma^2$ is independent of $\theta$.) For $m>0$,
if the coupling is weak at short length scales, then it flows to
larger values at larger length scales. For $\theta\neq\pi$ (mod
$2\pi$), the coupling becomes large, the U($n+m|n$) symmetry is
restored, and the theory is massive. However, a transition is
expected at $\theta=\pi$ (mod $2\pi$). For $m>2$, this transition
is believed to be first order, while it is second order for $m\leq
2$ \cite{affleck}. In the latter case, the system with
$\theta=\pi$ flows to a conformally-invariant fixed-point theory.
At the fixed point, a change in $\theta$ is a relevant
perturbation that makes the theory massive. Our exact results
describe these critical theories for $m\leq 2$.

For $m=2$, the RG flow is the same as for the familiar O(3) sigma
model. For $n=0$, the $\theta=\pi$ critical theory is the SU(2)
level 1 Wess-Zumino-Witten (WZW) model \cite{affprl}. For $n>0$, a
supersymmetric extension of this familiar system is obtained. For
$m=1$, we argue that the critical theory is related to the
critical CFT of percolation.

For $m=0$, the beta function is exactly zero (see Sec.\
\ref{perc}), and the sigma model exhibits a line of fixed points,
with continuously-varying scaling dimensions for general $n>1$. We
argue that these dimensions do not depend on $\theta$. We find an
exact description of the spectrum in one theory with the same
symmetry, and argue that it represents one special value of the
coupling $g_\sigma^2$. This theory describes ``dense polymers''
(there are also variant dense polymer theories with OSp($2n|2n$)
supersymmetry). These points will be discussed further in Sec.\
\ref{perc}; an important role is played by the fact that for
$n=1$, this theory is just non-interacting massless scalar
fermions (with central charge $c=-2$). This case has similarities
with the sigma model with a supergroup manifold SL($n|n$)/GL(1) as
target space in Ref.\ \cite{berk,bersh} (for a careful discussion
of the target manifold, see Ref.\ \cite{zirn99}).

The other nonlinear sigma model has target space $S^{2n+m-1|2n}$;
it is a supersymmetric extension of the usual O($m$)-vector model.
As coordinates, we use a real scalar field, $$\phi\equiv(\phi_1,
\ldots,\phi_{2n+m},\psi_1,\ldots,\psi_{2n}),$$ (the $2n+m$
components $\phi_a$ are commuting, the remaining $2n$ components
$\psi_\alpha$ are anticommuting), and
\begin{equation}
\phi\cdot\phi'=\sum_a\phi_a\phi'_a+\sum_\alpha{\cal
J}_{\alpha\beta}\psi_\alpha\psi'_\beta
\end{equation}
is the osp($2n+m|2n$)-invariant bilinear form [${\cal
J}_{\alpha\beta}$ is a nondegenerate symplectic form, for which we
can take the standard form, consisting of diagonal blocks
$\left(\begin{array}{rr}0&1\\-1&0\end{array}\right)$]. In the
nonlinear sigma model, the fields are restricted to obey the
constraint
\begin{equation}
\phi\cdot\phi=1,
\end{equation}
which defines the unit supersphere, $S^{2n+m-1|2n}$. The
Lagrangian density of the sigma model is
\begin{equation}
{\cal L}=\frac{1}{2g_\sigma^2}\partial_\mu \phi\cdot
\partial_\mu\phi.\label{sigmaS}
\end{equation}
The perturbative beta function for the coupling $g_\sigma^2$ in
this model is (again, it is independent of $n$ to all orders)
\begin{equation}
\beta(g_\sigma^2)=(m-2)g_\sigma^4+O(g_\sigma^6). \label{ombet}
\end{equation}
For $m>2$, the RG flow is towards strong coupling at large length
scales, and it is expected that the symmetry is restored, and no
transition occurs (for $n=0$, $m\geq 2$, a broken symmetry phase
is not allowed in 2D, and no phase with power-law correlations is
known in this model for $n=0$, $m>2$). For $m<2$, zero coupling is
an attractive fixed point, and one then expects a transition to
separate this regime from strong coupling. Our exact results
describe this critical point. In the case $m=1$, $n=0$, it is best
to use a lattice version of the model, as will be described in
Sec.\ \ref{susy}, and then this is simply the 2D Ising phase
transition. For $m=0$, this critical point is used to describe
dilute polymers (self-avoiding walks).

For $m=2$, the theory is more interesting, as the perturbative
beta function for $g_\sigma^2$ vanishes to all orders, similar to
the case of the ${\bf CP}^{n-1|n}$ model. For the action as given,
we again expect that the beta function vanishes exactly, and there
is a line of fixed point theories with continuously-varying
scaling dimensions. However, the lattice version of the model,
discussed in Sec. \ref{susy}, is the usual $XY$ model for $n=0$,
and a transition occurs due to vortex excitations \cite{kt}, which
are not accounted for in the action (\ref{sigmaS}). The
supersymmetric extension should have excitations that correspond
to vortices, that can appear as perturbations in the action, and
which become relevant above a critical value of the (renormalized)
coupling $g_\sigma^2$. Then, as in the Kosterlitz-Thouless (KT)
theory of the 2D $XY$ model \cite{kt}, there will be a low
temperature (small $g_\sigma^2$) phase with power-law correlations
with continuosly-varying exponents, and a high-temperature (large
$g_\sigma^2$) phase that is massive, separated from each other by
a critical point. It is for this critical theory that we obtain
the exact spectrum.

The critical CFTs in the $S^{2n+m-1|2n}$ models can also be
reached starting from Landau-Ginzburg or $\phi^4$ theories, in
which the field $\phi$, unconstrained this time, has Lagrangian
density
\begin{equation}
{\cal L}=\frac{1}{2}\partial_\mu\phi\cdot\partial_\mu\phi
+\frac{1}{2}r\phi\cdot\phi+\frac{1}{8}\lambda(\phi\cdot\phi)^2.
\label{parso}
\end{equation}
Such a supersymmetric approach to dilute polymers (the $m=0$ case)
was proposed by Parisi and Sourlas \cite{parsou} (this works in
any spacetime dimension). In these theories, a transition can
occur when the renormalized mass-squared $r$ changes sign, again
only if $m\leq2$. At the critical point, $\lambda$ flows away from
zero in any dimension less than four, and reaches a nontrivial
fixed point. These critical points should be the same CFTs as in
the nonlinear sigma models of the same symmetry. The ``low
temperature'' ($r<0$) regime of these models should correspond to
the weakly-coupled regime of the $S^{2n+m-1|2n}$ sigma models.

We now turn to the lattice models we use, which originated as
descriptions of percolation, polymers, and some other problems in
statistical mechanics. The standard approach to percolation,
polymers (self-avoiding walks) and related 2D problems leads to a
CFT description of the critical phenomena involving a single
scalar field with a background charge, and has been well-studied.
It is less widely appreciated that there are representations of
the original lattice models using supersymmetry \cite{parsou}. The
cancellation between fermion and boson states in loops
(percolation cluster boundaries, or polymer loops) in graphical
expansions can be used to weight these loops by 1 (for
percolation) or 0 (for polymers), in place of the usual $Q\to 1$
limit in percolation (based on the $Q$-state Potts model) or $m
\to 0$ limit in polymers (based on the O($m$)-vector model). In
the cases of percolation \cite{glr} and of dense polymers, we will
show that the supersymmetric model is a vertex model (or in an
anisotropic limit, a quantum spin chain) and there is a
corresponding nonlinear sigma model, the ${\bf CP}^{n+m-1|n}$
model at $m=1$ (percolation), $0$ (dense polymers). For dilute
polymers, the corresponding nonlinear sigma model is the
$S^{2n+m-1|2n}$ model at $m=0$. The consequence of the existence
of supersymmetric models for these problems is that there should
be versions of the CFT descriptions that exhibit the same global
supersymmetries.

Our results for the percolation case apply directly to the
critical theory of the spin quantum Hall (SQH) transition in
noninteracting fermions with quenched disorder, for which some
exact exponents have been found previously by a mapping of a
lattice model onto percolation \cite{glr}. While much has been
learned about percolation and polymers without the use of the
supersymmetric models, in the random fermion problems such as the
SQH and IQH transitions the supersymmetry is virtually all we have
to guide us in seeking the critical theories. Therefore, a fuller
understanding of CFTs with the appropriate type of supersymmetry
should aid our understanding of the latter problems. Our results
constitute another step towards the understanding of previously
unknown CFTs that have superalgebras as symmetries, and central
charge $0$, $-2$ or other values. An important part of our results
is that these theories are not rational CFTs, even though all the
conformal weights are rational numbers, because the spectra do not
consist of a finite number of representations of a chiral algebra
of integer spin fields. In particular, the chiral algebras do not
contain affine Lie superalgebras, nor do the theories appear to be
related to such superalgebras in any simple way. While the
presence of symmetries acting locally on the degrees of freedom
does imply the existence of local currents, these do not generate
an affine Lie superalgebra because they are not purely right- or
left-moving (i.e., not chiral---a fact that will manifest itself
in logarithms in their correlators), and so cannot be part of any
chiral algebra. We anticipate that such features may be common in
the presently unknown CFTs for the random fermion problems.

We note that there is other previous work on applications of sigma
models with supergroup manifolds as the target space, in which the
CFTs are also not current algebras, though closely related to
them. These theories are related to other random fermion problems,
the so-called chiral ensembles
\cite{gweg,gade,hs,altsi,zirn99,gll}. There is a proposal to apply
one of these theories, using the work in Refs.\ \cite{berk,bersh},
to the IQH transition \cite{zirn99}. Certain other random fermion
problems can be tackled fruitfully by current-algebra methods
\cite{lfsg,ntw,mcw,ntw2,ctt,lb,awwl,bckt}.

We now describe our method, in brief overview. We first discuss
vertex models with supersymmetry, and their relation to nonlinear
sigma models and Landau-Ginzburg theories. Special cases of these
models describe some properties of  percolation and polymers. The
models can be defined with periodic boundary conditions on, say, a
rectangle with sides $l_1$, $l_2$. They can be viewed as 1D
quantum systems in the usual way, say by constructing a transfer
matrix $T=e^{-H}$, and then $Z={\rm STr}\, e^{-Hl_2}$ is the
partition function, which (by construction) is $Z=1$ for
percolation and dilute polymers, and $Z=0$ for dense polymers. The
supertrace, denoted $\rm STr$, is defined as ${\rm
STr}\,(\cdots)={\rm Tr}\,[(-1)^F\cdots]$, where $F=0$, $1$ is the
grading of states; in many models, $F$ is the fermion number. The
partition function $Z$ is 1 or 0 in these special cases because of
cancellations between states associated with the same energy
eigenvalue (i.e.\ eigenvalue of $H$). To obtain the energy
eigenvalues, we will calculate $Z_{\rm mod}={\rm Tr}\,e^{-Hl_2}$.
This means modifying the boundary condition in the 2 (imaginary
time) direction so as to insert an additional factor $(-1)^F$. For
the original periodic boundary conditions, all fields are periodic
around either fundamental cycle of the torus. In the modified
partition function, the boundary condition is modified (twisted)
such that odd (fermionic) fields are antiperiodic along the 2
(time) direction, periodic along the 1 (space) direction, while
even (bosonic) fields remain periodic. By varying the aspect ratio
of the system, the energies and momenta, and their multiplicities,
can be read off from the modified partition function. [We note
that to incorporate supersymmetry as an internal symmetry, it is
convenient to use a space of states that includes some that have
negative norm \cite{read,glr} (more pedantically, it is the
norm-squared that is negative). This is connected with the
violation of the spin-statistics theorem in the continuum field
theories. This allows $H$ to be Hermitian with respect to the
indefinite inner product, but nonetheless because of the negative
norms, viewed as a matrix $H$ is not diagonalizable \cite{km}. It
can be brought to Jordan normal form, and then the diagonal
elements of $e^{-Hl_2}$ are of the form $e^{-El_2}$ with
multiplicities given by the sum of the sizes of the Jordan blocks
that contain $E$. These Jordan blocks are related to the fact that
these CFTs are logarithmic \cite{gurarie}. We will not obtain any
information about the Jordan block structure, and will simply
think of energy eigenvalues $E$ with multiplicities.]

By evaluating the transfer matrix products, the vertex models
become loop models, that is sums over configurations of
nonintersecting loops, with various weights. In the supersymmetry
point of view, each loop has states in the defining representation
of the superalgebra running around the loop, and the state on a
loop is chosen independently of those on other loops. In
particular, for periodic boundary conditions each loop, whether or
not it winds the cycles of the torus, is given a weight which is
the supertrace (denoted ${\rm str}$) of $1$ in the defining
representation of the superalgebra, which is set up to be ${\rm
str}\,1=m$ in general, and thus 1 (for percolation), 0 (for
polymers). The effect of the modified boundary condition is that a
loop that winds along the $2$-direction an odd number of times
loses the cancellation of even and odd states, and is given a
factor which is the trace (not the supertrace) of $1$ in the
defining representation, while loops that wrap an even number of
times are given unchanged weights.

The loop models can in turn be represented using a scalar field
with a background charge, with certain boundary condition effects
(see Ref.\ \cite{dfsz} and references therein). We will show that
these can be modified to include the differing factor for each
loop that winds the $2$ direction an odd number of times. After
performing a Poisson resummation, we obtain the modified partition
function in the continuum limit, which gives the spectrum of each
CFT. For low-lying levels, the multiplicities are dimensions of
representations of the superalgebra, and this should continue to
hold for higher levels, where the multiplicities are positive
integers.

In Section \ref{susy} we describe explicitly the supersymmetric
vertex models we use, show how they can be mapped onto loop models
(special cases of which describe percolationa and polymers), and
discuss general aspects of the critical phenomena further. In
Sec.\ \ref{coulomb} we give the construction of the modified
partition function for a general loop model. In Sec.\
\ref{results} this is applied to the sigma (or Landau-Ginzburg)
models that correspond to percolation, dilute and dense polymers,
and also some other models. Appendix \ref{resum} gives details of
the derivation of the modified partition function. Appendix
\ref{finite} contains results on finite size systems with free
boundary condition in the space direction. These give the spectrum
of the boundary CFT for percolation, whereas the rest of the paper
is about the bulk CFTs. For this case, we show that the states
enumerated by the modified partition function are complete. Some
partial results on completeness of states in finite size with
periodic boundary conditions in the space direction are also
given.

\section{Supersymmetric models}
\label{susy}

In this Section, we describe the supersymmetric lattice models for
percolation and polymers. We also explain the relation to the
field theories introduced in Sec.\ \ref{intro}, and give some
further field-theoretic arguments.

%%%%%%%%%%%%%%%%%%%%%%%%%%%%%%%%%%%%%%%%%%%%%%%%%%%%%%%%%%%%
\subsection{Supersymmetric vertex models and ${\bf CP}^{n+m-1|n}$ sigma models}
\label{perc}

A supersymmetric representation of the standard model of bond
percolation in 2D was constructed in Ref.\ \cite{glr}. Some
salient points are reviewed here (some further details can be
found in Appendix \ref{finite}). We also describe the mapping to
the corresponding ${\bf CP}^{n+m-1|n}$ sigma models, and their use
in understanding the physics of the lattice models.

The standard model of 2D percolation is bond percolation on the
square lattice. Clusters can be represented alternatively by their
boundaries, which are configurations of space-filling,
non-intersecting (and non-self-intersecting) loops on edges of the
square lattice (the loops are allowed to touch, without
intersecting, at the vertices). The latter square lattice is the
medial graph of that on which the bonds are placed in percolation.
This mapping has been pictured, and the weights given, many times
(see e.g.\ \cite{bax,nienrev,dfsz,glr}), so we will omit the
details, except for the important point that there is a global
factor of $1$ for each loop, in addition to local weights that are
independent of the global properties. The model can be formulated
on a simply-connected portion of the plane, or on a cylinder, or
on a torus, that is with periodic boundary conditions in two
directions. The global factor per loop applies to loops that wind
around the nontrivial cycles of the cylinder or torus, as well as
to those that are homotopic to a point. Because percolation (or
Potts) clusters admit a two-coloring with opposite colors on the
two sides of any segment of a loop, the loops that wind around the
system must do so in pairs, and because they do not intersect, all
the loops that wind the system must be homotopic to one another.
The partition function is the sum over all distinct configurations
of such loops, with equal weight, such that $Z=1$.

A popular representation for the percolating clusters uses a
$Q$-state Potts model on the underlying graph. This yields a loop
model with the same local weights but a global factor $\sqrt{Q}$
per loop \cite{bax}, so percolation corresponds to the limit
$Q\to1$ (when the model is formulated on the torus, there is a
slight difference between the loop model and the Potts model
\cite{dfsz}). Affleck \cite{affleck2} showed that the loop model
can be reproduced using a representation of the transfer matrix in
a space with $m$ states per site, with $Q=m^2$. The transfer
matrix has SU($m$) symmetry.

We now describe the supersymmetric vertex model, which is closely
related to that in Ref.\ \cite{affleck2}, and a special case of
which represents percolation \cite{glr}. The transfer matrix acts
in a space that is a ${\bf Z}_2$-graded tensor product (denoted
$\otimes$) of ${\bf Z}_2$-graded vector spaces at each site; these
spaces have graded dimension $n+m|n$, with $n+m$, $n\geq0$ (we
will also refer to the total dimension of spaces, which here would
be $2n+m$). This space is represented using boson and fermion
oscillators, with constraints. We consider sites labeled $i=0$,
\ldots, $2l_1-1$, with periodic boundary condition. For $i$ even
we have boson operators $b_i^a$, $b_{ia}^\dagger$,
$[b_i^a,b_{jb}^\dagger]=\delta_{ij}\delta_b^a$ ($a$, $b=1$,
\ldots, $n+m$), and fermion operators $f_i^\alpha$,
$f_{i\alpha}^\dagger$,
$\{f_i^\alpha,f_{j\beta}^\dagger\}=\delta_{ij}\delta_\beta^\alpha$
($\alpha$, $\beta=1$, \ldots, $n$). For $i$ odd, we have similarly
boson operators $\overline{b}_{ia}$, $\overline{b}_i^{a\dagger}$,
$[\overline{b}_{ia},\overline{b}_j^{b\dagger}]=\delta_{ij}\delta^b_a$
($a$, $b=1$, \ldots, $n+m$), and fermion operators
$\overline{f}_{i\alpha}$, $\overline{f}_i^{\alpha\dagger}$,
$\{\overline{f}_{i\alpha},\overline{f}_j^{\beta\dagger}\}=
-\delta_{ij}\delta^\beta_\alpha$ ($\alpha$, $\beta=1$, \ldots,
$n$). Notice the minus sign in the last anticommutator; since our
convention is that the $\dagger$ stands for the adjoint, this
minus sign implies that the norms of any two states that are
mapped onto each other by the action of a single
$\overline{f}_{i\alpha}$ or $\overline{f}_i^{\alpha\dagger}$ have
opposite signs, and the ``Hilbert'' space has an indefinite inner
product.

The supersymmetry generators are the bilinear forms
$b_{ia}^\dagger b_i^b$, $f_{i\alpha}^\dagger f_i^\beta$,
$b_{ia}^\dagger f_i^\beta$, $f_{i\alpha}^\dagger b_i^b$ for $i$
even, and correspondingly $-\overline{b}_i^{b\dagger}
\overline{b}_{ia}$, $\overline{f}_i^{\beta\dagger}
\overline{f}_{i\alpha}$, $\overline{f}_i^{\beta\dagger}
\overline{b}_{ia}$, $\overline{b}_i^{b\dagger}
\overline{f}_{i\alpha}$ for $i$ odd, which for each $i$ have the
same (anti-)commutators as those for $i$ even. Under the
transformations generated by these operators, $b_{ia}^\dagger$,
$f_{i\alpha}^\dagger$ ($i$ even) transform as the fundamental
(defining) representation $V$ of gl($n+m|n$),
$\overline{b}_i^{a\dagger}$, $\overline{f}_i^{\alpha\dagger}$ ($i$
odd) as the dual fundamental $V^\ast$ (which differs from the
conjugate of the fundamental, due to the negative norms). We
always work in the subspace of states that obey the constraints
\begin{eqnarray}
b_{ia}^\dagger b_i^a+f_{i\alpha}^\dagger f_i^\alpha &=& 1 \quad
(i\hbox{ even}),\label{constr1}\\ \overline{b}_i^{a\dagger}
\overline{b}_{ia}-\overline{f}_i^{\alpha\dagger}
\overline{f}_{i\alpha} &=& 1 \quad (i\hbox{ odd})\label{constr2}
\end{eqnarray}
(we use the summation convention for repeated indices of types $a$
or $\alpha$). These specify that there is just one particle,
either a boson or a fermion, at each site, and so we have the
graded tensor product of alternating irreducible representations
$V$, $V^\ast$ as desired. In the spaces $V^\ast$ on the odd sites,
the odd states (those with fermion number
$-\overline{f}_i^{\alpha\dagger} \overline{f}_{i\alpha}$ equal to
one) have negative norm.

A note on the superalgebras is in order (these remarks are
well-known, and can also be found in e.g. Ref.\ \cite{bersh}). The
above bilinears generate the superalgebra gl($n+m|n$) in the
defining representation $V$; they span the Hermitian matrices on
$V$. One generator, denoted $\cal E$, and given by ${\cal E}=
b_{ia}^\dagger b_i^a+f_{i\alpha}^\dagger f_i^\alpha$ (or the
identity matrix) in the defining representation, commutes with the
others. The generators which correspond to the Hermitian matrices
with vanishing supertrace (in $V$), form a subsuperalgebra,
denoted sl($n+m|n$). For $m\neq0$, $\cal E$ has nonzero
supertrace, and is eliminated by restricting to sl($n+m|n$). In
these cases, sl($n+m|n$) is a simple superalgebra. But for $m=0$,
$\cal E$ has supertrace 0, and generates an ideal in sl($n|n$).
The quotient superalgebra, denoted psl($n|n$), can be obtained by
taking the quotient of sl($n|n$) by the ideal. For $n>1$,
psl($n|n$) is simple. The spaces $V$, $V^\ast$ are, strictly
speaking, not representations of psl($n|n$) because $\cal E$ is
nonzero in these spaces (${\cal E}=+1$ in $V$, $-1$ in $V^\ast$),
but the tensor products of equal numbers of factors $V$ and
$V^\ast$, as we use in our models, are. We also note that we use
lower case letters for the superalgebras (which are always over
${\bf C}$), while upper case letters such as U($n+m|n$) denote a
corresponding supergroup, with the real form of the underlying Lie
group specified---in this example, it is U$(n+m)\times $U$(n)$.
The real form is determined by the unitary transformations with
respect to the inner products in the spaces in our models.

The invariant transfer matrix is constructed as follows. First we
note that for any two sites $i$ (even), $j$ (odd), the
combinations
\begin{equation}
b_i^a \overline{b}_{ja} + f_i^\alpha\overline{f}_{j\alpha},\quad
\overline{b}_j^{a\dagger}b_{ia}^\dagger +
\overline{f}_j^{\alpha\dagger}f_{i\alpha}^\dagger \end{equation}
are invariant under gl($n+m|n$), thanks to our use of the dual
$V^\ast$ of $V$. Then the transfer matrix acting on sites $i$,
$i+1$ with $i$ even is
\begin{eqnarray}
T_i&=&(1-p_A)+p_A(\overline{b}_{i+1}^{a\dagger}b_{ia}^\dagger +
\overline{f}_{i+1}^{\alpha\dagger}f_{i\alpha}^\dagger)\nonumber\\
&&\times(b_i^a \overline{b}_{i+1,a} +
f_i^\alpha\overline{f}_{i+1,\alpha}),
\end{eqnarray}
while for $i$ odd
\begin{eqnarray}
T_i&=&p_B+(1-p_B)(\overline{b}_i^{a\dagger}b_{i+1,a}^\dagger +
\overline{f}_i^{\alpha\dagger}f_{i+1,\alpha}^\dagger)\nonumber\\
&&\times(b_{i+1}^a \overline{b}_{ia} +
f_{i+1}^\alpha\overline{f}_{i\alpha}).
\end{eqnarray}
The complete transfer matrix is
\begin{equation}
T\equiv T_1 T_3\cdots T_{2l_1-1}T_0T_2\cdots T_{2l_1-2},
\end{equation}
and $i=2l_1$ is interpreted as $i=0$. Then $Z={\rm STr}\,T^{l_2}$.
We usually assume $0\leq p_S\leq 1$, where $S=A$, $B$. By taking
either of the two terms in $T_i$ for each vertex in the graph, the
expansion in space-filling loops mentioned above is obtained
\cite{glr}, with factors $p_S$, $1-p_S$ for each vertex, and a
factor $(n+m)-m={\rm str}\, 1=m$ for each loop (in all the models
considered in this paper, we will consider only $m$, $n$ integer,
with $n+m$, $n\geq0$). The latter is equal to the supertrace in
the fundamental representation (denoted $\rm str$), of 1, because
the evaluation of contributions for each loop can be viewed in
terms of states in $V$ flowing around the loop. This holds true
for topologically nontrivial loops as well as for loops homotopic
to a point. The transfer matrix $T$ on a chain with periodic
boundary conditions, as considered here, is translationally
invariant (under translation by two sites), as well as
supersymmetric. We note that these constructions give
representations of the Temperley-Lieb algebra (at certain
parameter values determined only by $m$) in the various graded
vector spaces, as discussed further in App.\ \ref{finite}. The
special case $m=1$ gives percolation ($Z=1$ for $m=1$). This
version has the advantage of being mathematically well-defined,
because there is a mathematically-meaningful, nontrivial transfer
matrix at $m=1$, unlike the case of the SU($m$) model
\cite{affleck2} which requires that a poorly-defined limit $m\to1$
be taken in order to obtain any quantity of interest. For $m=1$,
the parameters $p_A$ and $p_B$ are the probabilities of links of
the square lattice being occupied by bonds in the percolation
problem, with $A$ as the horizontal, and $B$ as the vertical
links. For all $m$, the system is isotropic when $p_A=p_B$, while
the transition occurs when $p_A=1-p_B$.

In the anisotropic (time- or 2-continuum) limit $p_A\to 0$ with
$p_A/(1-p_B)$ fixed, we can write $T\simeq
e^{-p_A^{1/2}(1-p_B)^{1/2} H}$, where the Hamiltonian $H$ is
\begin{eqnarray}
H &=&-\epsilon\sum_{i\,{\rm even}}
(\overline{b}_{i+1}^{a\dagger}b_{ia}^\dagger +
\overline{f}_{i+1}^{\alpha\dagger}f_{i\alpha}^\dagger)(b_i^a
\overline{b}_{i+1,a} +
f_i^\alpha\overline{f}_{i+1,\alpha})\nonumber\\
&&\mbox{}-\epsilon^{-1}\sum_{i\,{\rm
odd}}(\overline{b}_i^{a\dagger}b_{i+1,a}^\dagger +
\overline{f}_i^{\alpha\dagger}f_{i+1,\alpha}^\dagger)(b_{i+1}^a
\overline{b}_{ia} +
f_{i+1}^\alpha\overline{f}_{i\alpha})\nonumber\\
&&\mbox{}+(\epsilon+\epsilon^{-1})l_1,\label{susyham}
\end{eqnarray}
and $\epsilon=\sqrt{p_A/(1-p_B)}$. This Hamiltonian, like the
transfer matrix above, is invariant under sl($n+m|n$) (or
psl($n|n$) for $m=0$) for all parameter values $p_A$, $p_B$. This
Hamiltonian can be viewed as that of an antiferromagnetic
superspin chain, built from the generators listed above times
those on the neighboring site, contracted using the invariant
bilinear form on the superalgebra, up to constant terms. When
$\epsilon\neq 1$, the two types of terms in $H$ have different
coefficients, and this perturbation (``staggering the coupling'')
is a relevant perturbation that takes the system off the
transition point (see e.g. Ref.\ \cite{glr}). For $m=1$, the
ground state energy of $H$ is zero, which follows from an argument
similar to that in Ref.\ \cite{km}, or from the facts that the
partition function is $1$, and that there are no energies less
than zero (which is less obvious). The preceding family of
supersymmetric vertex models, or the corresponding superspin chain
Hamiltonians, are the sl($n+m|n$)-invariant models for which we
find exact results in the following.

For such vertex models or spin chains, there is a corresponding
continuum quantum field theory, which is a nonlinear sigma model
with target space the symmetry supergroup [here U($n+m|n$)],
modulo the isotropy supergroup of the highest weight state (see
e.g.\ Refs.\ \cite{affleck,rs1} for related, but
non-supersymmetric, examples, and Refs.\ \cite{z,grl} for
supersymmetric examples in random fermion problems). For the
present cases we obtain U($n+m|n$)/U($1)\times$U($n+m-1|n) \cong
{\bf CP}^{n+m-1|n}$, a supersymmetric version of complex
projective space. This mapping can be controlled by generalizing
the spin chain model by replacing the fundamental by a ``higher
spin'' (larger) irreducible representation (irrep), represented by
the size of the highest weight $\cal S$, $2{\cal S}$ integer (with
${\cal S}=1/2$ for the fundamental), and its dual on alternate
sites. For an appropriate family of such irreps, such that the
isotropy supergroup is the same, the sigma model can be obtained
with a bare coupling constant (at the scale of the lattice
spacing) $g_\sigma^2\sim 1/{\cal S}$, and the mapping is exact at
large $\cal S$. In the construction above, this can be done by
replacing 1 on the right hand side of Eqs.\ (\ref{constr1}),
(\ref{constr2}) by the integer $2{\cal S}$, and using the same
Hamiltonian (\ref{susyham}), up to an additive constant (the
transfer matrix of the vertex model would need to be modified).
The action of the nonlinear sigma model obtained this way is as in
Eq.\ (\ref{cpnlag}), where the bare value of $\theta$ is
$\theta=2\pi {\cal S}+O(\epsilon-1)$; note that $\theta$ can be
varied by staggering the couplings \cite{affleck}. In fact, as
$\epsilon$ varies from $0$ to $\infty$, $\theta$ varies from $0$
to $4\pi{\cal S}$, and so passes through $\pi$ (mod $2\pi$)
$2{\cal S}$ times; for $2{\cal S}$ odd, one such point occurs at
$\epsilon=1$ for symmetry reasons.

A great deal of insight into the vertex models or spin chains of
interest, with ${\cal S}=1/2$, can now be gained by analyzing the
sigma models, and vice versa. When the beta function Eq.\
(\ref{umbet}) for $g_\sigma^2$ is positive at weak coupling, it is
expected that the same fixed point CFT is reached in the universal
large length-scale behavior of the spin chain models for all odd
values of $2{\cal S}$, when $\epsilon=1$ ($\theta=\pi$) (more
generally, also at all the $2{\cal S}$ critical values of
$\epsilon$, for all $\cal S$). In particular, for $m=1$, this CFT,
to which the ${\bf CP}^{n-1|n}$ sigma model at $\theta=\pi$ flows,
is related to the critical CFT of percolation. We should point out
that problems like percolation are defined geometrically, and that
it is possible to define many scaling dimensions, or correlation
functions, for geometric or topological properties of clusters. On
the other hand, we are dealing here with concrete models, and for
these the spectra of states and corresponding operators in CFT are
precisely defined, and may not include all possible operators that
yield geometric properties of clusters. The model we are using
emphasizes the properties of the loops, which are the boundaries
(hulls) of percolation clusters \cite{glr}.

As argued in Sec.\ \ref{intro}, for $m=2$, the RG flow in the
sigma model is the same as for the familiar O(3) sigma model. For
$n=0$, the spin chains are the familiar Heisenberg spin chains,
and for $2{\cal S}$ odd, the critical theory is the SU(2) level 1
Wess-Zumino-Witten (WZW) model \cite{affprl}. For $n>0$, a
supersymmetric extension of this familiar system is obtained. We
will return to results for this model in Sec.\ \ref{othsusy}
below.

Also, the case $m=0$ (and ${\cal S}=1/2$) can serve as a model for
dense polymers. The general set-up for polymers is discussed more
fully in Section \ref{poly}. Here we just specify that dense
polymers are obtained when one or more long polymer loops are
forced into contact. Models of dense polymers involve
space-filling non-intersecting loops, with a global factor $m=0$
for each loop. One such model, for loops on the square lattice, is
obtained from the psl($n|n$) invariant vertex model. The structure
of this model implies that it contains information on exponents
for only {\em even} numbers of polymers winding around the torus.
The partition function of this model vanishes, as long as the
boundary conditions are psl($n|n$)-invariant, as they are here.
The case $n=1$ is a free-fermion model \cite{ivash}. Define
fermion operators $F_i^\dagger=f_{i1}^\dagger b_i^1$ ($i$ even),
$\overline{F}_i=\overline{b}_i^{1\dagger}\overline{f}_{i1}$ ($i$
odd); the space of states is just a Fock space, without
constraints, in these variables. These operators flip the
superspins, and are related directly to the odd generators of the
superalgebra. The Hamiltonian in terms of $F$, $\overline{F}$ is
noninteracting, and can easily be solved exactly; indeed, for
$m=0$, $n=1$, this can be done for all $\cal S$ and $\epsilon$.
The result is free massless scalar complex, or ``symplectic'',
fermions \cite{parsou,gurarie,ivash}, with Lagrangian density in
the continuum limit ($\eta$, $\eta^\dagger$ are anticommuting
complex fields):
\begin{equation}
{\cal
L}=\frac{1}{2g_\sigma^2}\partial_\mu\eta^\dagger\partial_\mu\eta,
\label{sympfer}
\end{equation}
where (for each value of $\epsilon$) $g_\sigma^2$ is precisely
proportional to $1/2{\cal S}$. By construction, the free fermions
are the Goldstone modes (spin waves) of the spontaneously-broken
supersymmetry of the antiferromagnetic chain. The odd elements of
the superalgebra act on $\eta$ as $\eta\mapsto \eta+\xi$, $\xi$ a
complex constant, so no other terms with zero or two derivatives
are possible in the action. The spin-wave theory of the
antiferromagnet is exact in this case.

In the case of general $n$, the naive continuum (or large $\cal
S$) limit in this $m=0$ case is the sigma model with target space
${\bf CP}^{n-1|n}$. In this model, the perturbative beta function
vanishes identically. This can be seen either from direct
calculations, which have been done to at least four-loop order
\cite{wegner}, or from an argument similar to that in Ref.\
\cite{berk}: for $n=1$, the sigma model reduces to the massless
free fermion theory (\ref{sympfer}) (the sigma model target space
for $n=1$ has dimension $0|1$ over the complex numbers), and
further the theta term becomes trivial in this case. Thus, for all
sigma model couplings $g_\sigma^2>0$, the theory for $n=1$ is
non-interacting. The free-fermion theory is conformal, with
Virasoro central charge $c=-2$; $\theta$ is clearly a redundant
perturbation in this case, as it does not appear in the action (a
similar argument appeared in Ref.\ \cite{bsz}). By the same
argument as before, the conformal invariance with $c=-2$ should
hold for all $n$, and also for all $g_\sigma^2$ and $\theta$,
though the action is no longer non-interacting in general. Thus
the beta function is also identically zero non-perturbatively. In
general, the scaling dimensions will vary with the coupling
$g_\sigma^2$, so changing $g_\sigma^2$ is an exactly marginal
perturbation, though for $n=1$ the coupling can be scaled away, so
there is no dependence on the coupling in the exponents related to
those multiplets of operators that survive at $n=1$. Hence for
$n=1$, the exactly-marginal perturbation that changes $g_\sigma^2$
is redundant. The dense-polymer problem (the ${\cal S}=1/2$ case)
apparently corresponds to only one special value of $g_\sigma^2$.
We also expect that, at the special value of $g_\sigma^2$,
$\theta$ remains redundant for all $n$, because the vertex model
remains critical when staggering is present, as we know from the
$n=1$ case, and in the modified Coulomb gas derived below, this
constrains all the parameters, so staggering does not change any
scaling dimensions even for $n>1$ (whereas we cannot change
$g_\sigma^2$ within the vertex model at fixed ${\cal S}=1/2$, and
this leaves open the possibility that scaling dimensions, other
than those of operators that occur for $n=1$, change with this
coupling). It is likely that $\theta$ is redundant for all
$g_\sigma^2$ and all $n$.

Finally, the cases $m<0$ of the sl($n+m|n$) (or percolation-type)
models may also be of physical interest. For sufficiently weak
coupling, the flow of $g_\sigma^2$ is towards zero at large length
scales, for all $\theta$. There could be a transition at some
finite $g_\sigma^2$. When $n+m=1$, the action consists only of
interacting scalar fermions, and there can be no theta term. Then
using similar arguments as in the dense polymer case, the critical
theory will be independent of $\theta$ for all $n\geq 1-m$. The
exact results presented below may describe these transition
points, at least for $m=-2$, $-1$.

%%%%%%%%%%%%%%%%%%%%%%%%%%%%%%%%%%%%%%%%%%%%%%%%%%%%%%%%%%
\subsection{Polymers and $S^{2n+m-1|2n}$ sigma models}
\label{poly}

Polymers are modeled as self-avoiding and non-intersecting random
walks. There is a standard method in which this is represented
using a scalar field theory with $m$ component scalar fields, and
the limit $m\to0$ in which closed loops disappear yields the
statistics of polymers \cite{degennes}. Such a model is
O($m$)-invariant. Parisi and Sourlas introduced a supersymmetric
version, for which we gave the Lagrangian already in Eq.\
(\ref{parso}). The critical point of that theory for $m=0$
represents ``dilute'' polymers.

The lattice model usually used in obtaining exact results for 2D
polymers is a strongly-coupled version of the field theory,
obtained by truncating the high-temperature expansion of a lattice
model \cite{dmns}. A supersymmetric version of this model can be
constructed with OSp($2n+m|2n$) supersymmetry, as follows: We use
the honeycomb graph, which can be constructed by tiling the torus
with regular hexagons. The partition function is
\begin{equation}
Z=\int\prod_{i}\frac{d\phi_i}{(2\pi)^{n+m/2}}e^{-\sum_i\phi_i\cdot\phi_i/2}
\prod_{<ij>}(1+K\phi_i\cdot\phi_j).\label{Ommod}
\end{equation}
Here the $2n+m|2n$ component field $\phi_i$ at each site (vertex)
$i$ of the honeycomb graph has components as in Sec.\ \ref{intro},
and transforms in the fundamental (or vector) representation $V'$
of OSp($2n+m|2n$) [where the real form of the underlying Lie group
is O($2n+m$) $\times$ Sp($2n,{\bf R}$)]. $d\phi_i$ is the standard
OSp-invariant measure on $\phi_i$. Expanding the product over
distinct links $<ij>$ of the graph, and performing the
integrations over the fields, the partition function is given by
\begin{equation}
Z=\sum_{\rm loops}K^L m^{\#{\rm loops}},
\end{equation}
where the sum is over all configurations of any number of
non-space-filling non-intersecting closed loops on the honeycomb
graph, $L$ is the total number of edges of the graph occupied (at
most once) by part of a loop, and $\#{\rm loops}$ is the number of
loops. Notice the global factor ${\rm str}\,1=2n+m-2n=m$ per
closed loop, whether or not the loop is topologically nontrivial
[the supertrace $\rm str$ here is in the vector representation
$V'$]. In general, the critical point of the model occurs at
$K_c=1/\sqrt{2+\sqrt{2-m}}$ for $-2\leq m \leq 2$ \cite{nien}. For
$m=0$, $Z=1$, and at the critical value $K_c$ of $K$ already known
from the $n=0$ case, this yields the theory of dilute polymers.

The same loop model can be obtained from a similar model in which
the field is restricted to obey $\phi_i\cdot\phi_i=1$ for all $i$,
making this a lattice version of the $S^{2n+m-1|2n}$ sigma model.
The partition function of such a model would be
\begin{equation}
Z=\int\prod_{i}d[\phi_i]\,e^{-J\sum_{<ij>}\phi_i\cdot\phi_j/T},
\label{Omsigmod}
\end{equation}
where $d[\phi_i]$ is the normalized invariant measure on the
supersphere, $J$ is a constant, and $T$ is temperature. This
partition function can be shown to be independent of $n$, with no
approximation. The high-temperature (strong-coupling) expansion of
this model is obtained by expanding the exponential in powers of
$J/T$. It can be shown to be independent of $n$ to all orders. In
general the expansion involves diagrams with any number of lines
(with associated factor $J/T$ per line) on each link of the graph,
with a condition that an even number of lines end at each vertex.
Each line represents the vector representation propagating along
the links. We truncate the expansion of the exponential to the
first two terms \cite{dmns}, and then the graphical expansion is
the same as in the model above, with $K\propto J/T$. Either of
these constructions can be used as the supersymmetric version of
the O($m$) model. [For $m=1$, $n=0$, this model accurately
represents the (untruncated) Ising model on the honeycomb graph,
with $K=\tanh J/T$.] Finally, either model can be represented by a
transfer matrix on the tensor product of complex
$2n+m+1|2n$-dimensional spaces ${\bf C}\oplus V'$ (where $V'$ has
been complexified), one for each vertical link in a horizontal row
of the graph; the states represent either the empty (singlet) or
the occupied (vector) state.
%In $V'$, there is a natural basis in
%which exactly half the odd basis states have negative norm, so the
%unitary symmetry group on the odd subspace is the noncompact group
%Sp($2n,{\bf R}$), as required.
The partition function is then the supertrace of this transfer
matrix to the power $l_2$. One then expects the critical theory of
any of these models to be the same as in those discussed in the
introduction. We remark that the Lie superalgebra-invariant
approach to polymers is different from the $N=2$ superconformal
approach proposed in \cite{sal92}, although relations between the
two are not excluded.

Apart from the $m=0$ polymer case, the OSp($2n+m|2n$)-invariant
models (\ref{Ommod}) with $m\neq 0$ may also be of interest. The
cases $m>2$ show no critical phenomena, but the cases $m=1$, $2$
exhibit the standard Ising and KT transitions for $n=0$. Again,
for $n>0$ we obtain supersymmetric extensions which have
transitions at which the CFTs include the exponents and
correlators of the usual theories. In spite of having the same
supersymmetry, these models do not have the same representation
structure as the random-bond Ising model \cite{grl}. In Sec.\
\ref{othsusy} below, we also discuss some results for the case
$m=-2$.

In the O($m$) models (\ref{Ommod}), there is also a large $K$,
``low $T$,'' regime. For all $K$ in the range $K_c<K<\infty$, the
system flows to the same fixed point CFT (which varies with $m$).
For $m=2$, this coincides with the ``critical'' branch, discussed
above, but is a distinct CFT for $m<2$ \cite{nien}. For $m=0$,
this yields a model of dense polymers, which in this case has no
restriction on the numbers of loops that can wind the two
directions on the torus \cite{desclo}. We will consider this
briefly below. Naively, or based on the Landau-Ginzburg theory
(\ref{parso}), the low $T$ region of the O($m$) or
OSp($2n+m|2n$)-invariant models would be expected to be a broken
symmetry regime of the model (\ref{parso}), described by the
weakly coupled $S^{2n+m-1|2n}$ sigma model, see Eq.\
(\ref{sigmaS}). The perturbative beta function (\ref{ombet}) for
the coupling $g_\sigma^2$ in this model is such that for $m<2$
zero coupling is an attractive fixed point at large distances
(which tends to confirms the existence of a transition in this
range of $m$). However, this noninteracting behavior is at odds
with the conformally-invariant behavior of the O($m$) models,
which map to loop models, that we study here, so apparently the
latter do {\em not} describe this regime of the $S^{2n+m-1|2n}$
sigma models. Instead, the low $T$ O($m$)/loop models are more
closely related to the sl($n+m|n$)-invariant models, as we will
see. This distinction is most likely due to the truncation of the
high-$T$ expansions, as discussed above, which does not change the
universality class in the high-$T$ and critical regions, which are
the same as in the continuum theory (\ref{parso}), but could
change the low-$T$ behavior. In terms of the physics of polymers,
what we call dense polymers is the $m=0$ model in which dense
loops are strictly non-intersecting, and the resulting CFT has
$c=-2$. If we adopt instead de Gennes' model of weak repulsions,
as in Lagrangian (\ref{parso}), then intersections are suppressed,
but allowed, and it seems that this ``dense'' phase is attracted
to the non-interacting fixed point at $g_\sigma^2=0$ in the
$S^{2n-1|2n}$ sigma model, a theory of free scalar bosons and
fermions with central charge $c=-1$. (This distinction of the two
models relies on the topology of loops in 2D, so may not hold in
higher dimensions.) We expect that the introduction of a small
amplitude for allowing crossings in the former dense polymer model
produces a crossover to the latter fixed point---the apparent
violation of Zamolodchikov's $c$-theorem being possible, because
the theories in question are not unitary.

% This phase does not seem to coincide with the usual dense
%polymers in two dimensions. For instance, it has $c=-1$ while for
%the latter $c=-2$. This discrepancy may have its origin in the
%very strong constraints imposed by self avoidance in 2D, that may
%not be captured by the general formalism of PS in generic
%dimensionality.

In the Introduction we have already briefly discussed the model
with partition function (\ref{Omsigmod}) for $m=2$. For $n=0$, it
is the XY model, and the transition is understood in terms of
vortices \cite{kt}, which are characterized by integers, since the
fundamental homotopy group $\pi_1(S^1)={\bf Z}$. On the other
hand, for $n>0$, $\pi_1(S^{2n+1|2n}) =\pi_1(S^{2n+1})=0$, and
there are no topologically stable vortices in the continuum
$S^{2n+1|2n}$ sigma model. But we still believe that the partition
function of the lattice model on the torus with periodic boundary
conditions is the same for all $n$, and similarly for the
continuum model (\ref{sigmaS}) in which the fields are assumed
continuous, so there are no vortices. At weak coupling, we can
attempt a semiclassical calculation within the continuum sigma
model (\ref{sigmaS}). For $n=0$, one expands around the saddle
point where $\phi$ is constant, and a set of saddles where $\phi$
winds in one or both directions on the torus. [In fact, such
partition functions for winding numbers $M'$, $M$ appear in a
different context in eq.\ (\ref{partfn1}) below.] For $n>0$, the
natural saddle points that correspond to winding configurations
for $n=0$ are those where the field winds (as a function of
position going around a cycle on the torus) around a great circle
of the sphere, which break the symmetry down to osp($2n|2n$).
These saddle points are unstable, but the multiplicity of the
unstable modes is $2n|2n$, and so the negative eigenvalue cancels
in the formal one-loop fluctuation determinant. Summing over such
saddle points and working to one-loop order, we can reproduce the
partition function of the continuum $S^{1|0}$ model. Thus it is
very plausible that these configurations are associated with the
winding modes for $n=0$, which correspond in CFT to operators that
insert vortices.

%%%%%%%%%%%%%%%%%%%%%%%%%%%%%%%%%%%%%%%%%%%%%%%%%%%%%%%%%%%%%%%
\subsection{Modifying boundary conditions}

We outlined in the Introduction how we would modify the boundary
condition in the 2 direction so as to evaluate $Z_{\rm mod}={\rm
Tr}\,e^{-Hl_2}$ instead of $Z={\rm STr}\,e^{-Hl_2}$. Note that
this modified boundary condition breaks the supersymmetry. More
generally, we will evaluate $Z_{\rm mod}={\rm Tr}\,e^{i{\cal
P}\tau_Rl_1-Hl_2}$, where ${\cal P}$ is the momentum, and $\tau_R$
is a free real parameter. The insertion of the translation by
$\tau_Rl_1$ in the 1 direction makes the torus non-rectangular.

It is not difficult to see that the effect of the modified
boundary condition is that, for each loop that winds in the 2
direction (crosses the 1 axis), which would previously have picked
up the factor ${\rm str}\,1$, now picks up ${\rm
str}\,(-1)^{n_2f}$, where $n_2$ is the number of times it cuts the
1-axis, and $f=0$ or $1$ represents the grading of states in the
fundamental: for the sl($n+m|n$)- (respectively, osp($2n+m|2n$)-)
invariant models, $f=0$ for the first $n+m$ (resp., $2n+m$) basis
states in $V$ (resp., $V'$), $f=1$ for the remaining $n$ (resp.,
$2n$). Hence, loops that wind the 2 direction an even number of
times are given the factor $m$ as usual, while those that wind an
odd number of times are given $2n+m$ (resp., $4n+m$).

In discussing the spectrum, we will often refer to operators as
well as to states. The models we are studying are all compact,
that is there is a finite number of states per site in the chain
on which the transfer matrix acts (in the naive continuum field
theories, the target spaces are all compact, or effectively so as
excursions to large field values are suppressed). In such systems,
one expects that all operators that are strictly local in terms of
these variables (i.e., operators whose correlators with local
functions of these variables are single-valued) correspond to
states in the CFT, so the spectrum of one is the spectrum of the
other. There may also be meaningful operators that are not local,
such as twist fields or disorder operators in the Ising model.
These are not expected to correspond to states. Examples in the
models we study include the Potts spin operator, for which there
is a nonlocal construction in the supersymmetric vertex model
\cite{glr}, operators studied by Duplantier \cite{duplan} and by
Cardy \cite{cardy}, and probably the $0$-hull ($0$-leg) operator
for percolation (polymers) \cite{sd,sal86,sal92}.

The $n=1$ case of percolation is related to the $n=1$ case of the
model for the spin QH transition \cite{glr}. In general, the naive
continuum theory for the spin QH effect is the sigma model with
target space OSp($2n|2n$)/U($n|n$), where the real form of the
even part of osp this time is SO*($2n$)$\times$Sp($2n$) (class C
in Ref.\ \cite{az}). This target space is noncompact for $n>1$,
but compact and the same as ${\bf CP}^{1|1}$ for $n=1$. In the
noncompact cases, states lie in infinite-dimensional
representations, and the spectrum is expected to be continuous. On
the other hand, the natural local operators (at least, those that
are local functions of the sigma model fields) transform in
finite-dimensional representations, and the general argument above
implies that the operators for $n=1$ are a subset of those for
$n>1$. Thus our results give us information about the operators in
general, whose spectrum will be discrete. Similar statements about
the spectra of states and operators apply to the other noncompact
sigma models, as typically arise in the random fermion problems
including the IQH case \cite{az}.

%%%%%%%%%%%%%%%%%%%%%%%%%%%%%%%%%%%%%%%%%%%%%%%%%%%%%%%%%%%%%%%%%
\section{Modified Coulomb gas partition functions of loop models}
\label{coulomb}

In Section \ref{susy} we saw that various supersymmetric models
can be mapped onto loop models, where the global weight factor
associated with each loop is a supertrace in the fundamental
representation of the appropriate superalgebra. In this Section,
we show how the modified partition functions of general loop
models can be evaluated at the critical point, in the continuum
limit, with periodic boundary conditions (modified, for the 2
direction). We first set up a mapping of loop models onto a
Coulomb gas partition function following Ref.\ \cite{dfsz}, with
modifications. It is convenient to begin with the polymer-type
situation (the osp($2n+m|2n$)-invariant models of Sec.\
\ref{poly}), in which any number of loops can wind either
direction. Afterwards, we indicate the modification for the
percolation-type models (those in Sec.\ \ref{perc}), in which
these numbers must both be even.

The starting point for mapping loop models on a lattice onto
scalar field theories is to represent the configurations of loops
as configurations of a dual ``height model'' that assigns a real
``height'' variable to each plaquette of the lattice. The loops
are here viewed as having two possible orientations (represented
by an arrow on the loop) assigned to them, which will be summed
over. The heights associated to a given configuration of directed
loops are defined, up to addition of a constant, so that on
crossing a loop from left to right (looking along the direction of
the arrow on the loop), the height $\varphi$ changes by
$\varphi_0$, a constant. Locally, the heights are well-defined,
because of the absence of any sources or sinks for the arrows on
the loops. On the torus, loops that wind around the two cycles can
lead to heights that are not single valued, but are single-valued
modulo $\varphi_0$. Suppose the height changes by
$\delta_1\varphi=M\varphi_0$ in the 1 direction, and by
$\delta_2\varphi=M'\varphi_0$ in the 2 direction (where $1$ and
$2$ refer to a choice of fundamental cycles on the torus, which is
not necessarily rectangular now). Then, because the loops are
non-intersecting, it can be shown \cite{dfsz} that this
corresponds to at least $|M\wedge M'|$ distinct loops, each of
which wind the 2 direction $|M/(M\wedge M')|$ times, and the 1
direction $|M'/(M\wedge M')|$ times (disregarding the orientation
of each loop), plus any number of loops that are homotopic to a
point. Here $m\wedge n$ denotes the highest common factor of two
integers, with $m\wedge 0=m$ for all $m\geq0$.

Now each loop that is homotopic to a point is given a global
factor $m=2\cos \varphi_0e_0$ by the sum over the two possible
orientations, and the same factor is desired for each
homotopically non-trivial loop in the unmodified partition
function. For our modified partition function, this can be simply
modified to $2\cosh \alpha$ for each loop that winds the $2$
direction an odd number of times. The arguments of Ref.\
\cite{dfsz} can now be easily modified to show that, in the
continuum limit taken at appropriate values for the local weight
factors (discussed in Sec.\ \ref{susy}), the modified partition
function can be written as
\begin{eqnarray}
\lefteqn{Z_{\rm
mod}[g,\varphi_0,e_0,\alpha]=}\nonumber\\&&\qquad\sum_{M',M\in{\bf
Z}}Z_{M',M}(gf^2)w(M,d;f,e_0,\alpha),\label{partfn}\\
\lefteqn{w(M,d;f,e_0,\alpha)=}\nonumber\\&& \qquad\cos (2\pi
dfe_0)\;\delta_{M/d\equiv0}+\cosh (d\alpha) \;\delta_{M/d\equiv1}.
\end{eqnarray}
where $d=M'\wedge M$, $f=\varphi_0/(2\pi)$, and in the Kronecker
$\delta$'s the congruences are modulo $2$. The notation is
somewhat redundant, since the parameters $g$, $f$, $e_0$ enter
only in the combinations $gf^2$ and $fe_0$, but it is conventional
and we will keep it. The continuum scalar field partition function
is given by
\begin{eqnarray}
Z_{M',M}(g)&=& Z_0(g)\exp\left\{-\pi g
[M'^2+M^2(\tau_R^2+\tau_I^2)\right.\nonumber\\ &&\mbox{}\left.-2
MM'\tau_R]/\tau_I\right\}, \label{partfn1}
\end{eqnarray}
and
\begin{equation} Z_0(g)=
\frac{\sqrt{g}}{\tau_I^{1/2}\eta(q)\eta(\overline{q})}.
\end{equation}
$\tau=\tau_R+i\tau_I=\omega_2/\omega_1$ is the complex aspect
ratio (modular parameter) of the torus ($\tau_I=l_2/l_1>0$),
$q=e^{2\pi i\tau}$, and
\begin{equation}
\eta(q)=q^{1/24}\prod_{n=1}^\infty(1-q^n)
\end{equation}
is the Dedekind function. We note that the Virasoro central charge
of the continuum CFT is
\begin{equation}
c=1-6e_0^2/g,
\end{equation}
and the coupling $g$ for the scalar field depends on the details
of the local weights in general. The expression (\ref{partfn}) is
clearly not modular invariant. However, the underlying unmodified
partition function, in which the weight is always the cosine
factor, is modular invariant by construction, because of the sum
over all winding numbers that it contains. $Z_{\rm mod}$ does not
have to be modular invariant, as it is simply a generating
function for the spectrum.

Next we perform a Poisson resummation on the $M'$ variable. For
each $M\neq 0$, this has to be done by grouping the terms into
sums of $Z_{M',M}(gf^2)$ over $M'$ a multiple of some positive
integer (possibly 1). As in Ref.\ \cite{dfsz}, the general
structure of the result is
\begin{eqnarray}
\lefteqn{Z_{\rm mod}[g,2\pi f, e_0,\alpha]=}&&\nonumber\\
&&\quad\frac{1}{\eta(q)\eta(\overline{q})}\left\{ \sum_{P}
q^{\Delta_{e_0+P/f,0}(g)}\overline{q}^{\overline{\Delta}_{e_0+P/f,0}(g)}
\right.\nonumber\\
&&\quad\;\mbox{}+\!\!\sum_{\stackrel{\scriptstyle M>0,N>0,P:}
{N|M,P\wedge N=1}}\Lambda_{\rm mod}(M,N;f,e_0,\alpha)\nonumber\\
%&&\left.\mbox{}\sum_{P\in{\bf Z}:P\wedge N=1}
&&\quad\;\;\mbox{}\left.\vphantom{\sum_{P}}\times
q^{\Delta_{P/(Nf),Mf}(g)}\overline{q}^{\overline{\Delta}_{P/(Nf),Mf}(g)}
\right\}.
\label{partfn2}
\end{eqnarray}
The summations are over the integers, except for restrictions
shown explicitly; note that, for any integers $N$, $M$, $N|M$
means $M$ is divisible by $N$. Also
\begin{eqnarray}
\Delta_{e,m}(g)&=&{1\over 4} \left({e\over
\sqrt{g}}+m\sqrt{g}\right)^{\!2},\nonumber\\
\overline{\Delta}_{e,m}(g)&=&{1\over 4}\left({e\over
\sqrt{g}}-m\sqrt{g}\right)^{\!2}, \label{Delt}
\end{eqnarray}
for any values of the electric charge $e$ and the magnetic charge
$m$. The coefficients $\Lambda_{\rm mod}(M,N;f,e_0,\alpha)$ are
derived in Appendix \ref{resum}, with the result (more compact
than that in Ref.\ \cite{dfsz})
\begin{eqnarray}
\lefteqn{\Lambda_{\rm mod}(M,N;f,e_0,\alpha)=}\nonumber\\
&&\quad2\sum_{d>0:d|M} \frac{\mu\left(\frac{N}{N\wedge
d}\right)\phi(M/d)}{M\phi\left(\frac{N}{N\wedge d}\right)}
w(M,d;f,e_0,\alpha).\label{Lamb}
\end{eqnarray}
In this expression, $\mu$ and $\phi$ are the M\"{o}bius and Euler
functions, respectively, the definitions of which are recalled in
App.\ \ref{resum}. The unmodified partition function $Z$ is given
by the same expressions but with $\Lambda$ in place of
$\Lambda_{\rm mod}$; $\Lambda$ differs from $\Lambda_{\rm mod}$ by
having $\cos 2\pi dfe_0$ in place of $w(M,d;f,e_0,\alpha)$,
irrespective of whether $M/d$ is even or odd.

In terms of CFT, the right- and left-moving conformal weights of
the states are found by expanding $Z_{\rm mod}$ in powers of $q$
and $\overline{q}$, and comparing with the form
\begin{equation}
Z_{\rm mod}=(q\overline{q})^{-c/24}{\rm Tr}
\,q^{L_0}\overline{q}^{\overline{L}_0};
\end{equation}
the right- (resp., left-) moving conformal weight $h$
($\overline{h}$) is the eigenvalue of $L_0$ ($\overline{L}_0$).
Alternatively, in terms of the Hamiltonian
$H=2\pi(L_0+\overline{L}_0)/l_1$ and momentum ${\cal P}
=2\pi(L_0-\overline{L}_0)/l_1$, the energy and momentum
eigenvalues (measured from the ground state values) are
$E=2\pi(h+\overline{h})/l_1$, ${\cal P}=2\pi(h-\overline{h})/l_1$.
The momentum is related to the conformal spin $s=h-\overline{h}$,
which is always an integer.

For the percolation-type, sl($n+m|n$)-invariant models, $M'$ and
$M$ are both even, hence so is $M'\wedge M$. Then we can move a
factor of two into $\varphi_0$ and $\alpha$, so now
$f=\varphi_0/\pi$, and the modified partition function is given by
$Z_{\rm mod}[g,2\varphi_0=2\pi f,e_0,2\alpha]$, as given in Eqs.\
(\ref{partfn2}), (\ref{Delt}), (\ref{Lamb}). We are now ready to
find the spectra of the various models discussed in Section
\ref{susy}.

%%%%%%%%%%%%%%%%%%%%%%%%%%%%%%%%%%%%%%%%%%%%%%%%%%%%%%%%%%%%%%%%
\section{Results}
\label{results}

In this Section we apply the formulas of the preceding Section to
the models in Section \ref{susy}, by recalling the values of the
parameters that enter in the continuum limit as in Ref.\
\cite{dfsz}. We begin with the percolation (${\bf CP}^{n|n}$) case
as it has direct application to the spin QH transition of recent
interest. We will refer to ``states'' or to ``operators''
interchangeably, where the operators are those that correspond to
the states. This does not rule out the possible interest of
operators that do not correspond to states, but they would be
nonlocal with respect to the variables used in the construction of
our models.

%%%%%%%%%%%%%%%%%%%%%%%%%%%%%%%%%%%%%%%%%%%%%%%%%%%%%%%%%%%%%%%
\subsection{${\bf CP}^{n|n}$ model at $\theta=\pi$, and percolation}
\label{percres}

To obtain the model of percolation hulls, or the ${\bf CP}^{n|n}$
sigma model at $\theta=\pi$, we have $m=2\cos(\pi e_0/2)=1$ (we
use $e_0=\pm 2/3$), $f=1/2$, $2\cosh\alpha=2n+1$, and we use the
result that in the continuum theory, $g=8/3$ \cite{dfsz}, and
hence $c=0$. By the remarks at the end of Sec.\ \ref{coulomb}, we
must calculate $Z_{\rm mod}[8/3,\pi,2/3,2\alpha]$. First we point
out some conformal weights of interest that appear in the
spectrum. If we ignore the contributions from expanding the
products in the Dedekind functions, which we view loosely as
giving ``descendants'', then the scalar (momentum or conformal
spin zero) states are either of purely electric type, with
dimensions
\begin{equation}
h=\overline{h}=\Delta_{e_0+2P,0}-{1\over 24}= {(3P+1)^2-1\over
24},
\end{equation}
or of purely magnetic type, with dimensions
\begin{equation}
h=\overline{h}=\Delta_{0,M/2}-{1\over 24}={4M^2-1\over 24}.
\end{equation}
The latter correspond to the $M$-hull operators ($M>0$) \cite{sd},
while the former are thermal \cite{dennijs}, and the sequences
overlap infinitely often. There are many other values at both zero
and nonzero conformal spin. For $M>0$, the general right-moving
conformal weight is given by
\begin{equation}
h={(3P/N+2M)^2-1\over 24},
\end{equation}
(plus positive integers in the case of descendants), and the
conformal spin is $s=PM/N$ (plus integers). This $h$ corresponds
to position $(-P/N,M)$ in the Kac table of $c=0$ Virasoro
representations. Since $N$ can be arbitrarily large for some $M$
(as long as $N|M$ and $N\wedge P=1$), many nondegenerate Virasoro
representations occur in the spectrum. These include some with $h$
in the range $-1/24<h<0$, however, even when $h$ or $\overline{h}$
is negative, $h+\overline{h}$ (or the excitation energy) is
positive. The multiplicities of these states are nonzero, since
the lowest weight $h$ at which a given $N$ occurs is when $N=M$
and $P$ is close to $-2M^2/3$ (such a state has conformal spin of
order $-2N^2/3$ for large $N$). Then the multiplicity is just
$\Lambda_{\rm mod}(N,N)$, which goes as $\sim(2\cosh 2N\alpha)/N$
for large $\alpha$ (or $n$). Thus there are infinitely many
equivalence classes of conformal weights under congruence modulo
1. This implies that, even though the conformal weights are all
rational numbers, the theory is not a rational CFT, since there
can be no chiral algebra (containing only chiral fields of integer
spin) under which the states fall into a {\em finite} number of
representations.

Next we give some results for the total multiplicities (including
descendants) of low-lying levels. The ground state
$h=\overline{h}=0$ is non-degenerate. The next-lowest levels are
the one-hull operators, with $h=\overline{h}=1/8$, which occur in
the two series mentioned above. The total multiplicity with which
these occur in $Z_{\rm mod}$ is
\begin{eqnarray}
D_1&=&1+\Lambda_{\rm mod}(1,1)=1+2\cosh 2\alpha\nonumber\\
&=&(2n+1)^2-1=D_{\rm adj},
\end{eqnarray}
the dimension of the adjoint representation of sl($n+1|n$), as
expected \cite{glr}. It is significant that this number is
positive, even, and can be identified as (in general, a sum of)
dimensions of representations of sl($n+1|n$) (it must be even,
since except for the ground state all states must cancel in pairs
when $Z$ is calculated). If the CFT underlying this system
contained right- and left-moving affine Lie superalgebra of
sl($n+1|n$) at some level, then the multiplicities of primary
fields would have to be products, or in the simplest case squares,
of dimensions of representations of sl($n+1|n$). Taking into
account the left-right symmetry of the system, that is obviously
not the case.

The next levels include the two-hull operators, with
$h=\overline{h}=5/8$. These occur for $M=0$, $P=1$, and $M=2$,
$P=0$, $N=1$. The total multiplicity is
\begin{eqnarray}
D_2&=&1+\Lambda_{\rm mod}(2,1)=1+\cos\pi
e_0+\cosh4\alpha\nonumber\\ &=& (2n+1)^4/2-2(2n+1)^2+3/2.
\end{eqnarray}
For $n=1$, this is $D_2=24$. This may possibly be a sum of
adjoints (each of dimension 8) and an indecomposable
representation of sl($2|1$) of dimension $8$ (discussed further in
Appendix \ref{finite}). The singlet operator in the indecomposable
can appear as a relevant ``thermal'' (non-symmetry breaking)
perturbation of the critical theory.

Among non-scalar operators, the most interesting are the spin-1
``currents'' $h=1$, $\overline{h}=0$.  They are obtained from (i)
$M=0$, as a descendant of the identity, (ii) $M=1$, $P=1$, $N=1$.
The total multiplicity is $1+\Lambda_{\rm mod}(1,1)=D_1=D_{\rm
adj}$. Thus they are precisely the currents of the global
sl($n+1|n$) supersymmetry, which must exist by Noether's theorem,
but which are apparently not chiral since if they were they would
generate an affine Lie superalgebra. Hence, they must be
logarithmic operators \cite{gurarie}: even though
$\overline{h}=0$, these operators have nontrival action on the
left-moving sector.

There is also a large multiplet of fields (one of them being the
stress energy tensor) with  $h=2$, $\overline{h}=0$: the total
multiplicity is $D_1+D_2$, which may be interpreted as descendants
of the currents, together with a ``primary'' multiplet of
dimension $D_2$. The latter is presumably again an indecomposable
representation, since the stress tensor, which is invariant under
sl($n+1|n$), must be paired with other fields so that $Z$ can be
$1$ \cite{gl}.
%The operators in the indecomposable multiplet should be chiral,
%like the stress tensor, since they can be mapped onto it by a
%global symmetry operation.

Finally, we mention the marginal operators of dimension
$(h,\overline{h})=(1,1)$. The total multiplicity turns out to be
$2D_1$. It is likely that these actually form two adjoints, and
that there is no invariant marginal operator. For $n>0$, this is
much less than would exist if there were right- and left-moving
chiral currents of weight 1, that would generate an affine Lie
superalgebra; in that case there would be at least $(D_{\rm
adj})^2$ weight $(1,1)$ states. Since there are not, we again
conclude that the currents are not chiral.

We should comment here that the weights $(1/3,1/3)$ do not appear
in the spectrum. These were proposed \cite{glr} as the weights of
a local spin-zero symmetry-breaking perturbation that had been
studied numerically \cite{kag}. Problems with this proposal were
pointed out previously \cite{cardy,gl}. The present results show
that no such local operator exists in the model in question, and
the operator in the superspin chain found in Ref.\ \cite{glr} must
have some other conformal weights in the long-distance limit. The
only spin-zero operators close by are the one- and two-hull
operators identified above. It was argued that the requisite
operator appears as the leading term in the operator product of
the one-hull operator with itself. Since the one-hull operator has
even parity \cite{glr}, it can appear in the product and now seems
to be the best candidate. This is consistent with the fact that
the operator in Ref.\ \cite{glr} is not a singlet, but part of a
multiplet, which however does not contain a singlet---just like
the one-hull operators in the adjoint representation. This would
predict that the exponent $\mu$ of Ref.\ \cite{kag} (see also
Ref.\ \cite{glr}) should be $\mu=2/(2-h-\overline{h})=8/7\simeq
1.14$, within $\simeq 25\%$ of the numerical value, $1.45$.

%%%%%%%%%%%%%%%%%%%%%%%%%%%%%%%%%%%%%%%%%%%%%%%%%%%%%%%%%%%%
\subsection{Critical $S^{2n-1|2n}$ model and polymers}
\label{polyres}

A similar analysis can be done for the dilute polymer model, that
is the critical point in the $S^{2n-1|2n}$ sigma model. In this
case, the parameters are $m=2\cos(\pi e_0)=0$ (we use $e_0=1/2$),
$f=1/2$, $2\cosh\alpha=4n$, $g=3/2$. The modified partition
function is $Z_{\rm mod}[3/2,\pi,1/2,\alpha]$. Many aspects of the
results are similar to the ${\bf CP}^{n|n}$ case, already
discussed: The theory is irrational, since it contains infinitely
many conformal weights that do not differ by integers. The weights
are
\begin{equation}
h={(8P/N+3M)^2-4 \over 96}
\end{equation}
plus non-negative integers for $M>0$, and $h=[(4P+1)^2-1]/24$ plus
non-negative integers for $M=0$. The spin zero operators include
some with the former weights with $P=0$ (the $M$-leg operators,
$M>0$ \cite{sal86}), and the latter weights which are those of
thermal operators \cite{carham,nien}.

Some of the low-lying states are as follows. The lowest excited
states are at $h=\overline{h}=5/96$, the 1-leg operators. Their
multiplicity is $D_1=\Lambda_{\rm mod}(1,1)=4n$. This is the
dimension of the defining (``vector'') representation of
osp($2n|2n$), as expected (and see a similar result in Ref.\
\cite{dfsz}). As in the discussion of percolation, this speaks
against identifying the CFT as containing an affine osp($2n|2n$)
Lie superalgebra. The next lowest spin-zero states are the
thermal, 2-leg operators, with $h=\overline{h}=1/3$, and
multiplicity $D_2=1+\Lambda_{\rm mod}(2,1)=8n^2$. It makes sense
to identify this representation of osp($2n|2n$) with the
supersymmetric second-rank tensors built from the graded tensor
product of the vector representation with itself, which has the
correct dimension (``supersymmetric'' here means they are
symmetric in the even-even, and antisymmetric in the odd-odd
components). It is indecomposable and contains a singlet. (For
$n=1$, we can use the isomorphism osp($2|2$) $\cong$ sl($2|1$),
and this indecomposable representation is the same as the
$8$-dimensional one mentioned in the last Subsection.) In a
Landau-Ginzburg view of the critical theory as an RG fixed point
in the theory (\ref{parso}), this singlet operator would be
$\phi\cdot\phi$, which at $\lambda=0$ clearly lies in a multiplet
of similar supersymmetric tensors, and this should be maintained
at finite $\lambda$ (and hence in the critical theory), because
the degeneracy of the multiplet is needed to maintain the
partition function $Z=1$ throughout the RG flow. (Similar
arguments regarding the multiplicities of energy levels at each
momentum in a finite size system can be given for the
sl($n+m|n$)-invariant models, using the sigma model at weak
coupling. Of course, such arguments cannot predict coincidences of
energies that occur in the critical theory, or at special values
of the coupling.) There are also other relevant spin-zero
($h=\overline{h}<1$) operators.

At nonzero spin, there are $8n^2$ currents with $h=1$,
$\overline{h}=0$. The currents have multiplicity $D_2$, which
equals the dimension $D_{\rm adj}$ of the adjoint representation
of osp($2n|2n$), as appropriate for the currents. The adjoint
representation consists of the super-antisymmetric second-rank
tensors, that are antisymmetric in the even-even part, and so on.
At weights $(h,\overline{h})=(2,0)$, there is a total of $D_1+D_2$
states. Even though the subset of $D_2$ states look formally like
descendants of the currents, we prefer to interpret them as
transforming in the supersymmetric representation, with the
singlet as the stress tensor \cite{gl}. This assignment would
again be expected from the Landau-Ginzburg point of view, where
the stress tensor has the form
$\partial\phi\cdot\partial\phi+\ldots$, and so is part of the
supersymmetric representation. This is clear at $\lambda=0$, and
should be maintained at finite $\lambda$, and hence in the
critical theory. However, the currents might be expected to have
Virasoro descendants at level 1, so our interpretation may have
some problems in CFT, if not in LG theory; this issue is
presumably connected with the logarithmic operators that are part
of the currents.

Finally, there are marginal operators at $(1,1)$, with
multiplicity $2D_2$. For $n>0$, this is again fewer than in
current algebra. This multiplet is most likely two adjoints, as in
percolation, so that there are no invariant marginal
perturbations.

%%%%%%%%%%%%%%%%%%%%%%%%%%%%%%%%%%%%%%%%%%%%%%%%%%%%%%%%%%%%%%%%
\subsection{${\bf CP}^{n-1|n}$ model and dense polymers}
\label{denspolres}

For dense polymers, we first discuss the model with psl($n|n$)
supersymmetry, then the osp($2n|2n$)-invariant model, which
contains additional scaling dimensions. For the former, the
modified partition function we require is $Z_{\rm mod}[g,2\pi
f,e_0,2\alpha]$, with $g=2$, $f=1/2$, $e_0=\pm1$,
$2\cosh\alpha=2n$. The central charge is $c=-2$.

For $n=1$, the expression for $Z_{\rm mod}$ simplifies
drastically. This is the case in which the lattice model is
essentially a psl($1|1$) spin chain, which is a free fermion
system that can be solved directly, so we expect to obtain the
modified partition function of symplectic fermions \cite{ivash}.
The expression Eq.\ (\ref{partfn}) for $Z_{\rm mod}$ becomes for
the percolation-type models
\begin{eqnarray}
\lefteqn{Z_{\rm mod}[g,2\varphi_0=2\pi
f,e_0,2\alpha]=}\nonumber\\&&\qquad\sum_{M',M\in{\bf
Z}}Z_{M',M}(gf^2)w(M,d;f,e_0,2\alpha),\label{partfn3}\\
\lefteqn{w(M,d;f,e_0,2\alpha)=}\nonumber\\&& \qquad\cos (2\pi
dfe_0)\;\delta_{M/d\equiv0}+\cosh (2d\alpha)
\;\delta_{M/d\equiv1},
\end{eqnarray}
where again $d=M'\wedge M$. Inserting the above values for dense
polymers and $n=1$, the Poisson resummation and use of the Jacobi
identity give
\begin{eqnarray} \lefteqn{Z_{\rm mod}[2,\pi,1,0]}&&\nonumber\\
&=&{1\over\eta(q)\eta(\overline{q})}\sum_{m \in 2{\bf Z},e\in {\bf
Z}+1/2} q^{(2e+m)^2/8}\overline{q}^{(2e-m)^2/8}\nonumber\\
&=&4\left|q^{1/12}\prod_{n=1}^\infty(1+q^n)^2\right|^2 \nonumber\\
&=&\hbox{det}(-{\cal D}_{\rm AP}),\label{zmoddens}
\end{eqnarray}
where ${\cal D}_{\rm AP}$ is the Laplacian on the torus, with
periodic boundary condition along the 1 cycle, and antiperiodic
along the 2 cycle. Thus this is indeed the free fermion partition
function. This conformal field theory contains the psl($1|1$)
affine Lie superalgebra in its chiral algebra (as remarked in a
similar context in Ref.\ \cite{sal92}); the right-moving currents
are $\partial \eta$, $\partial \eta^\dagger$.

For $n>1$, there are in general no cancellations, and the
situation is similar to the percolation and dilute polymer cases.
The values for the conformal weights, neglecting descendants, are
\begin{equation}
h={(2P/N+M)^2-1\over 8},
\end{equation}
for $M>0$, and $h=[(2P+1)^2-1]/8$ for $M=0$. In particular, for
spin zero, these give only the values $h=\overline{h}=(M^2-1)/8$
($M>0$) which are the dimensions of the $2M$-leg operators
\cite{sal92}. These are expected as our model contains only even
numbers of polymer loops winding on the torus, and in our model
local operators always conserve the number of polymer lines mod
$2$. All conformal weights are non-negative.

The ground state is degenerate, and its degeneracy is
$D_1=2+\Lambda_{\rm mod}(1,1)=4n^2$. This was expected, as a
natural generalization of the four-fold degeneracy of the
symplectic fermion ground state for $n=1$ \cite{kausch}, as in
both cases the states form the indecomposable adjoint
representation of gl($n|n$) (isomorphic to $V\otimes V^\ast$). The
singlet is the identity operator, and the multiplet represents the
2-leg operators. Notice that this multiplet cannot be decomposed
into representations of right- and left-moving psl($n|n$)
algebras, even for $n=1$. The next spin-zero states occur at
$h=\overline{h}=3/8$, the 4-leg operators. The multiplicity for
general $n$ is $D_2=\Lambda_{\rm mod}(2,1)=8n^2(n^2-1)$, so they
are absent for $n=1$. At nonzero spin, we find at
$(h,\overline{h})=(1,0)$ a total of $2D_1+D_2=8n^4$ states. These
are more than simply the $4n^2-2$ currents associated with
psl($n|n$) symmetry. At $(2,0)$ we find a multiplicity
$3D_1+D_2=4n^2(2n^2+1)$. The marginal operators, $h=1$,
$\overline{h}=1$, have total multiplicity $16n^2(4n^4-3n^2+2)/3$
(which is integer for all $n$). (All these multiplicities agree
with those for free fermions at $n=1$.) For $n=1$, the only
invariant marginal operator is redundant, corresponding to the
fact that in the free fermion theory, a change in coupling
$g_\sigma^2$ can be scaled away and has no effect. As explained in
Sec.\ \ref{perc}, if our theory describes a special point on the
${\bf CP}^{n-1|n}$ sigma model line of CFTs, an exactly marginal
(but not redundant) operator would be expected for $n>1$. There is
nothing to rule out the possibility that the invariant operator of
the $n=1$ case becomes such an operator for $n>1$, and the
identification is presumably correct.

An alternative model of dense polymers is the low $T$ regime of
the osp($2n|2n$)-invariant polymer-type model (\ref{Ommod}). The
continuum limit of this model is the same as for the psl($n|n$)
model, except that the restriction in the latter to even numbers
of loops $M'$, $M$ winding the torus is dropped, which means that
we use $f=1/4$, $\alpha$ (not $2\alpha$), and $2\cosh\alpha=4n$
for osp($2n|2n$) symmetry. Then the modified partition function we
need is $Z_{\rm mod}[2,\pi/2,\pm1,\alpha]$. As expected, in this
theory, $M$-leg operators \cite{sal92} with all $M>0$ occur, in
particular $M=1$, with conformal weight $h=\overline{h}=-3/32$
\cite{nien}. The latter has multiplicity $4n$, the dimension of
the vector representation. The identity and $2-$leg operators at
$h=\overline{h}=0$ have multiplicity $8n^2$ in this case, which we
presume is the indecomposable supersymmetric representation, with
the identity as the singlet. The structure of higher levels is
generally similar to the psl($n|n$) dense polymer model: one can
check that all conformal weights in the spectrum of the latter
still occur, but there are many additional weights, and the
multiplicities appear to reflect the osp($2n|2n$) symmetry, as in
the examples just given.

%%%%%%%%%%%%%%%%%%%%%%%%%%%%%%%%%%%%%%%%%%%%%%%%%%%%%%%%%%%%%%%
\subsection{Other supersymmetric models}
\label{othsusy}

Here we briefly consider some other models that arise by taking
$m\neq 0$, $1$ in the sl($n+m|n$)-invariant models, or $m\neq 0$
in the osp($2n+m|2n$)-invariant models.

One such model is obtained at $m=2$ in the supersymmetric vertex
model, or the ${\bf CP}^{n+1|n}$ sigma model at $\theta=\pi$. For
$n=0$, this is simply the SU(2)-invariant point of the six-vertex
model, or spin-1/2 Heisenberg spin-chain, without staggering of
the couplings. The critical theory is the SU(2) WZW model at level
1. For $n>0$, operators with the same conformal weights as those
in the latter theory will appear, and possibly others. We study
this using our modified partition function, with parameter values
$m=2$, so $e_0=0$, $f=1/2$, $2\cosh\alpha=2n+2$, and $g$ takes the
self-dual value $g=4$. The central charge $c=1$, and the conformal
weights are
\begin{equation}
h=\Delta=(P/N+M)^2/4,
\end{equation}
plus positive integers. For $n=0$, one checks (using M\"{o}bius
inversion again on $G$ in Eq.\ (\ref{Gform})) that our formulas
reduce to the standard Coulomb gas partition function; in this
case $\Lambda_{\rm mod}(M,N)=2\delta_{N,1}$. However, for $n>0$,
all values $N>1$ show up, and the structure is quite similar to
the ${\bf CP}^{n|n}$ case: the theory is not rational for any
choice of chiral algebra.

Some low-lying conformal weights are $(1/4,1/4)$, which has
multiplicity $2+\Lambda_{\rm mod}(1,1)=(2n+2)^2$. This is correct
for $n=0$, and for general $n$ is the square of the dimension of
the fundamental representation. However, since the fundamental of
sl($n+2|n$) is not self-dual, the WZW model for this symmetry
algebra should have twice this multiplicity for these states (thus
this WZW model does not go continuously to the limit $n=0$). This
is larger (by 1) than the dimension of the adjoint, which would be
the answer expected from the weak-coupling sigma model analysis
alluded to in Sec.\ \ref{polyres} (since the partition function
$Z>1$ in the present case, such isolated singlets are possible
above the ground state), but it is the minimal result that agrees
with the $n=0$ case. The currents, with weights $(1,0)$, have
multiplicity $1+\Lambda_{\rm mod}(1,1)=(2n+2)^2-1=D_{\rm adj}$
again in this case. The next lowest spin-zero states are at (1,1)
and have total multiplicity $3+\Lambda_{\rm mod}(2,1)+\Lambda_{\rm
mod}(1,1)=(2n+2)^4/2+1$. For $n=0$, this is $9=(D_{\rm adj})^2$,
but for $n>0$ is again less than $(D_{\rm adj})^2$. Even though
for $n=0$ this theory is precisely a WZW model, for $n>0$ it is
not, and the operators in it that correspond to those for $n=0$
must have operator products that involve those not present at
$n=0$, and that violate the current algebra structure.

It may be useful here to say a few words more formally about the
relation of the theories at different $n$, which we have mentioned
informally or by {\it ad hoc} arguments at various points in this
paper. The relevant set-up has been described already in Ref.\
\cite{bersh}, so we can be brief. The point is that a suitable
element ${\cal Q}=\sum_i{\cal Q}_i$, a sum of positive odd roots
${\cal Q}_i=\sum_{\alpha=1}^{n} f_{i\alpha}^\dagger
b_i^{2+\alpha}$ ($i$ even and the corresponding sl($n+2|n$)
generators for $i$ odd) in the superalgebra, obeys ${\cal Q}^2=0$,
and generates a correspondence of the even and odd states that we
would like to cancel. In the space of states, the cohomology space
${\rm Ker}{\cal Q}/{\rm Im}{\cal Q}$ (where Ker is the kernel, Im
is the image, of a map) can be shown explicitly to be isomorphic
to the $n=0$ Hilbert space. More generally, one can relate $n$ to
$0\leq n'<n$ in this way, and similar relations hold for other $m$
and for osp algebras. The algebras of operators on these spaces
are similarly related by taking the cohomology of the algebra in
the larger $n$ theory (where $\cal Q$ acts on operators by graded
commutation), both in finite size and in the continuum CFTs. In
particular, note that the chiral algebra of the cohomology can be
larger than the cohomology (which is chiral) of the chiral
algebra. In the ${\bf CP}^{n+1|n}$ case just discussed, the chiral
algebra of the cohomology is the SU($2)_1$ affine Lie algebra, but
this is not contained in the chiral algebra of the sl($n+2|n$)
invariant theory. The chiral algebra of the latter may be more
like a W-algebra, as in the theories in Ref.\ \cite{bersh}.
Similar arguments apply to the cancellation of integrations over
even and odd variables in the integrals that we have also used
earlier.

In the osp($2n+m|2n$)-invariant models, the cases $m=1$ and $m=2$
give supersymmetric extensions of the Ising and KT transitions,
respectively. Similar phenomena as for the SU(2) model just
discussed are expected here; they are {\em not} be described by
osp($2n+m|2n$) affine Lie superalgebras. In particular for $m=1$,
$n=0$, the Ising transition is related to a free Majorana fermion
field. The loop gas description reproduces the correct
modular-invariant partition function on the torus for the Ising
model \cite{dfsz}. For the critical point in the $S^{2n|2n}$ sigma
model with $n>0$, one might imagine that the theory becomes $2n+1$
free Majoranas, and $n$ beta-gamma (bosonic ghost) pairs, which is
a representation of osp($2n+1|2n$) current algebra at level 1
(using the normalization for the level that is standard for O($m$)
current algebra) \cite{gow}. However, that is not what occurs in
our model, and the theory is interacting, not free.

The other case $m=2$ in the osp($2n+m|2n$)-invariant, polymer-type
models is interesting. The partition function we find is of
similar structure as the others. This $S^{2n+1|2n}$ sigma model
possesses a line of fixed points with continuously varying scaling
dimensions (like the ${\bf CP}^{n-1|n}$ model, though here $c=1$),
and the KT point, at which we find the spectrum, is a special
point where all the dimensions become rational. The multiplicities
are quite similar to other cases. For example, the spin zero
states include the vector representation of multiplicity $4n+2$ at
$h=\overline{h}=1/16$, and a traceless supersymmetric tensor, of
multiplicity $(4n+2)^2/2$, at $h=\overline{h}=1/4$. The states at
$h=1$, $\overline{h}=0$ have multiplicity $(4n+2)^2/2-1$, the
dimension of the adjoint, so they can be identified with the
currents. There are $(4n+2)^4/4$ marginal operators at
$h=\overline{h}=1$. This coincides with $(D_{\rm adj}+1)^2$ in
this case, so is actually larger then $(D_{\rm adj})^2$. We expect
that this set includes fourth-rank supersymmetric tensors, and
others, and there should be just two invariant marginal operators.
One of the perturbations represents a change in the coupling
$g_\sigma^2$ in the $S^{2n+1|2n}$ sigma model, which in the
absence of the other perturbation is exactly marginal. By
perturbing the action by both these operators, it should be
possible to reproduce the Kosterlitz RG flow diagram near the KT
fixed point \cite{kt}.

The case $m=-2$ in the critical O($m$) ($S^{2n-3|2n}$) models is
also amusing. Here the Landau-Ginzburg theory (\ref{parso})
predicts that, for $n=1$, the theory is free massless symplectic
fermions \cite{parsou}, and $c=-2$. For $n=1$, we do indeed find
the same critical modified partition function as in eq.\
(\ref{zmoddens}). For $n>1$, we find that the set of conformal
weights $(h,\overline{h})$ is the same as in the psl($n|n$) (${\bf
CP}^{n-1|n}$) dense polymer model, but the multiplicities are
different, and reflect the dimensions of representations of
osp($2n-2|2n$). For example, at $h=\overline{h}=0$, there are $4n$
operators, which we interpret as two singlets plus the vector. One
invariant operator is the identity, the other is the relevant
perturbation that corresponds to $\phi\cdot\phi$ in the
Landau-Ginzburg picture, which for $n=1$ is just the fermion mass
term. At $(1,0)$, we find $8n^2-4n+2$ operators, which we
interpret as the $8n^2-8n+3$ currents in the adjoint, plus the
derivatives of the $(0,0)$ fields other than the identity. At the
next-lowest spin-zero weights, $h=\overline{h}=3/8$, we find
$8n(n-1)$ states, the dimension of the traceless supersymmetric
second-rank tensor of osp($2n-2|2n$). In the Ginzburg-Landau
picture, the trace of the multiplet of supersymmetric second-rank
tensors in $\phi$ is the invariant $\phi\cdot\phi$ identified at
$h=\overline{h}=0$; this is allowed by symmetry requirements. It
is clear that the theory for $n>1$ is not free scalar fields.

%%%%%%%%%%%%%%%%%%%%%%%%%%%%%%%%%%%%%%%%%%%%%%%%%%%%%%%%%%%%%
\section{Conclusion}
\label{conclusion}

To conclude, we have found a rich structure in the conformal field
theories of a number of critical points that arise in nonlinear
sigma models with target spaces ${\bf CP}^{n+m-1|n}$ and
$S^{2n+m-1|2n}$, and are related to models in statistical
mechanics such as percolation and polymers (critical and low $T$).
For $m$ an integer in the range $-2\leq m \leq 2$, these give 14
different series of theories, of which we explicitly mentioned
results for 8 (two further cases with $m=-2$ lie at $c=-\infty$,
so should be excluded). All the theories contain only rational
conformal weights, yet except in some exceptional cases at
particular $n$, for $n>0$ there are infinitely many fractional
conformal weights, and there can be no chiral algebra that will
organize them into a finite number of representations, so the
theories are not rational. None of the theories with $n>0$, except
the psl($1|1$) model, contains the affine Lie superalgebra
associated with its global supersymmetry in its chiral algebra, as
shown by the absence of commuting left and right actions of the
global symmetry on the multiplets of states, and other reasons. On
the other hand, the large multiplicities of the states point to an
enlarged global symmetry algebra of nonlocal charges, which is at
least as large as the commutant of the Temperley-Lieb algebra in
the lattice models (or some analogue in the osp-invariant cases).

There are many interesting open problems. It would be of great
interest to find the full symmetry structure of the operators, the
chiral algebra, and the operator product expansions of the fields.
With this information, it may be possible to extend the results to
the $n>1$ spin quantum Hall transition, and even to the integer
quantum Hall transition and other noncompact supersymmetric
nonlinear sigma models in general.

%%%%%%%%%%%%%%%%%%%%%%%%%%%%%%%%%%%%%%%%%%%%%%%%%%%%%%%%%%%%%

\acknowledgments

We are grateful to E. Witten for discussions of the ${\bf
CP}^{n-1|n}$ nonlinear sigma model, and to I.A. Gruzberg and A.
Ludwig for discussions and help with the references. This work was
supported by the NSF under grant no.\ DMR-98-18259 (N.R.), and by
the DOE (H.S.).

%%%%%%%%%%%%%%%%%%%%%%%%%%%%%%%%%%%%%%%%%%%%%%%%%%%%%%%%%%%%%%%%%%
\appendix

%%%%%%%%%%%%%%%%%%%%%%%%%%%%%%%%%%%%%%%%%%%%%%%%%%%%%%%%%%%%%%%
\section{Resummation of modified partition function}
\label{resum}

In this Appendix, we give the derivation of the result
(\ref{partfn2}), (\ref{Delt}), (\ref{Lamb}) from Eq.\
(\ref{partfn1}).

First, in the $M=0$ terms, $M/M'\wedge M$ is always even, and as
in Ref.\ \cite{dfsz} these terms can be resummed to give
\begin{equation}
\frac{1}{\eta(q)\eta(\overline{q})}\sum_{P}
q^{\Delta_{e_0+P/f,0}(g)}\overline{q}^{\overline{\Delta}_{e_0+P/f,0}(g)}\label{mequalzero}.
\end{equation}

In the $M\neq0$ terms, the terms $\pm|M|$ give the same result,
and we write them in terms of $M>0$, with a factor of $2$. Then
for each $M>0$, the sum can be written
\begin{equation}
Z_{\pm M}=2\sum_{d>0:d|M} w(M,d;f,e_0,\alpha)\sum_{M':M'\wedge
M=d}Z_{M',M}(gf^2).
\end{equation}
We need to find the sums
\begin{equation}
f(d)=\sum_{M':M'\wedge M=d}Z_{M',M}(gf^2)
\end{equation}
(in the following, $d$, $d'$, etc.\ will always stand for positive
divisors of $M$, i.e.\ $d|M$). On the other hand, sums over $M'$
of the following form can be Poisson-resummed:
\begin{eqnarray}
g(d)&\equiv&\sum_{M':d|M'}Z_{M',M}(gf^2)\nonumber\\
&=&\frac{1}{d\,\eta(q)\eta(\overline{q})}\sum_P
q^{\Delta_{P/(df),Mf}(g)}\overline{q}^{\overline{\Delta}_{P/(df),Mf}(g)}.
\end{eqnarray}
Clearly, we have
\begin{equation}
g(d)=\sum_{d':d|d'}f(d').
\end{equation}
Then one of the M\"{o}bius inversion formulas (pp.\ 234--239 in
Ref.\ \cite{hw}) can be used to obtain
\begin{equation}
f(d)=\sum_{d':d|d'}\mu(d'/d)g(d').
\end{equation}
The M\"{o}bius function $\mu$ is defined by $\mu(n)=(-1)^r$, if
$n$ is an integer that is a product $n=\prod_{i=1}^r p_i$ of $r$
{\em distinct} primes, $\mu(1)=1$, and $\mu(x)=0$ otherwise or if
$x$ is not an integer. The two central properties of $\mu(x)$ that
lead to the Mobius inversion formulas are \cite{vlw}
\begin{equation}
\sum_{d':d|d',d'|d''}\mu(d'/d)=\delta_{d,d''}, \label{mobius1}
\end{equation}
which we use here, and
\begin{equation}
\sum_{d':d|d',d'|d''}\mu(d''/d')=\delta_{d,d''}. \label{mobius2}
\end{equation}
These can be proved by noticing that for $d\neq d''$, there are
finitely many terms, and they cancel in pairs. Hence we have
\begin{eqnarray}
Z_{\pm M}
&=&\sum_{d'}G(M,d';f,e_0,\alpha)\frac{1}{\eta(q)\eta(\overline{q})}\nonumber\\
&&\mbox{}\times\sum_{P'}
q^{\Delta_{P'/(d'f),Mf}(g)}\overline{q}^{\overline{\Delta}_{P'/(d'f),Mf}(g)},
\end{eqnarray}
where
\begin{equation}
G(M,d';f,e_0,\alpha)=2\sum_{d}w(M,d;f,e_0,\alpha)\mu(d'/d)/d'.
\label{Gform}
\end{equation}

We now wish to group together the terms in which $P'$ and $d'$
have common factors. For each term in the sum, let $d''=P'\wedge
d'$, and $P'=Pd''$, $d'=Nd''$, and we can use $N>0$ as a summation
variable in place of $d''$. Then
\begin{eqnarray}
Z_{\pm M}
&=&\sum_{d'}G(M,d';f,e_0,\alpha)\frac{1}{\eta(q)\eta(\overline{q})}\nonumber\\
&&\mbox{}\times\sum_{\stackrel{\scriptstyle N>0,P:} {N|d',P\wedge
N=1}}q^{\Delta_{P/(Nf),Mf}(g)}\overline{q}^{\overline{\Delta}_{P/(Nf),Mf}(g)}\nonumber\\
&=&\sum_{\stackrel{\scriptstyle N>0,P:} {N|M,P\wedge
N=1}}\Lambda_{\rm mod}(M,N;f,e_0,\alpha)
\frac{1}{\eta(q)\eta(\overline{q})}\nonumber\\
&&\quad\mbox{}\times
q^{\Delta_{P/(Nf),Mf}(g)}\overline{q}^{\overline{\Delta}_{P/(Nf),Mf}(g)},
\end{eqnarray}
where
\begin{eqnarray}
\lefteqn{\Lambda_{\rm mod}(M,N;f,e_0,\alpha)=}&&\nonumber\\
&&2\sum_{d,d':N|d',d|d'}w(M,d;f,e_0,\alpha)\mu(d'/d)/d'.
\end{eqnarray}
Finally, the sum over $d'$ can be performed:
\begin{eqnarray}
\sum_{d':N|d'}\mu(d'/d)/d'&=&M^{-1}\sum_{t>0 :t|(M/N)}
t\,\mu(M/(dt))\nonumber\\ &=& c_{M/d}(M/N)/M,
\end{eqnarray}
where $t$ is an integer, and $c_n(m)$ is Ramanujan's sum (Ref.\
\cite{hw}, Thm.\ 271). By Thm.\ 272 of Ref.\ \cite{hw}, and some
elementary manipulation, we obtain
\begin{equation}
\sum_{d':N|d'}\frac{\mu(d'/d)}{d'}=\frac{\mu\left(\frac{N}{N\wedge
d}\right)\phi(M/d)}{M\phi\left(\frac{N}{N\wedge d}\right)},
\end{equation}
and Eq.\ (\ref{Lamb}) follows. Here $\phi(n)$ is Euler's totient
function, which is defined for positive integers $n$ as the number
of integers $m$ such that $1\leq m\leq n$ and $n\wedge m=1$, and
is given explicitly by
\begin{equation}
\phi(n)=n\prod_{p:p|n}(1-1/p),
\end{equation}
where the product is over primes $p$. This completes the
derivation of Eq.\ (\ref{partfn2}).

%%%%%%%%%%%%%%%%%%%%%%%%%%%%%%%%%%%%%%%%%%%%%%%%%%%%%%%%%

\section{Finite size and completeness of states}
\label{finite}

In this Appendix we supplement the main text by presenting results
for one case, the percolation model, in a finite size system with
free, rather than periodic, boundary conditions. In this model, we
can verify that our set of states is complete, that is, the number
of states of the supersymmetric chain coincides with the number of
states counted by the modified partition function. The exponents
found are the boundary hull exponents, and their descendants. The
continuum limit for free boundary conditions in other cases can be
found similarly. We also sketch a partial proof for completeness
of the states for periodic boundary conditions.

%We consider a model whose space of states is the graded tensor
%product ${\cal H}_{susy}=V\otimes V^\ast\otimes V\otimes
%V^\ast\cdots=(V\otimes V^\ast)^{l_1}$. The elementary transfer
%matrices act on neighboring spaces in the tensor product, and are
%of two types. Type $A$ acts on $V\otimes V^\ast$, and reads
%
%\begin{equation}
%T_A=t_A^2 P_. +(1-t_A^2) I\otimes\overline{I},
%\end{equation}
%
%where $I$ ($\overline{I}$) denotes the identity in $V$ ($V^\ast$),
%$P_.$ the projector onto the singlet representation. Type $B$ acts
%on $V^\ast\otimes V$ and reads
%
%\begin{equation}
%T_B=t_B^2 \tilde{P}_. +(1-t_B^2) \overline{I}\otimes I,
%\end{equation}
%
%where $\tilde{P}_.$ is again the projector onto the singlet, but
%this time in $V^\ast\otimes V$.  We chose a labeling of states
%such that $V$ (resp.\ $V^\ast$) is spanned by $|m\rangle$ (resp.\
%$|\overline{m}\rangle$), $m=0$, $1$, \ldots, $2n$. States with even
%labels $m$ are assumed bosonic, while the ones with odd labels are
%fermionic. The projectors $P_.$ then read
%
%\begin{eqnarray}
%\left(P_.\right)_{i\overline{j}}^{k\overline{l}}=\delta_{ij}
%\delta_{kl}(-)^l,\nonumber\\
%\left(\tilde{P}_.\right)_{\overline{i}j}^{\overline{k}l}= \delta_{ij}
%\delta_{kl}(-)^i.
%\end{eqnarray}
%
%with the obvious graphical representation given in figure 1.

We consider the system with free boundary conditions. We write the
operators $T_i$ defined in Sec.\ \ref{perc} as
\begin{eqnarray}
T_i&=&(1-p_A) + p_A e_i, \quad \hbox{($i$ even)}\nonumber\\
T_i&=&p_B+(1-p_B)e_i, \quad \hbox{($i$ odd)}.
\end{eqnarray}
Then $e_i$, $i=0$, \ldots, $2l_1-2$, satisfy the Temperley-Lieb
algebra relations
\begin{eqnarray}
e^2_i&=&me_i,\nonumber\\ e_i e_{i\pm 1}e_i&=&e_i,\nonumber\\
\left[e_i,e_j\right]&=&0,\quad j\neq i, i\pm 1\label{tlrel}
\end{eqnarray}
(in fact, these are satisfied by $e_i$ for all $i=0$, \ldots,
$2l_1-1$). One can also introduce a trace ${\rm Tr}_m$, which in
the supersymmetric construction is ${\rm Tr}_m={\rm STr}$, and
which obeys $m\,{\rm Tr}_m\,e_i W={\rm Tr}_m\,W$ ($i=0$, \ldots,
$2l_1-2$), where $W$ is any string of $e_j$'s with $j=0$, \ldots,
$i-1$. The relations (\ref{tlrel}), together with the
normalization for the trace ${\rm Tr}_m\, 1=m^{2l_1}$, are
sufficient to evaluate the trace ${\rm Tr}_m$ of any product of
$e_i$, with $i=0$, \ldots, $2l_1-2$, independent of the vector
space (or graded vector space) on which the Temperley-Lieb algebra
is represented.
%(Similar statements presumably hold in the
%periodic case, though care is needed with the relations involving
%${\rm Tr}_m$.)
Here we will consider only $m=1$.

We are mainly interested in the spectrum of the transfer matrix
obtained by multiplying the elementary vertex matrices along a
given row:
\begin{equation}
T=T_1 T_3\cdots T_{2l_1-3}T_0T_2\cdots T_{2l_1-2}.
\end{equation}
To obtain this spectrum, we consider the usual trace $\rm Tr$ of
the transfer matrix $T$ raised to the power $l_2$, that is, the
modified partition function of the model on an annulus. This
modified partition function can be expanded graphically. The
result is a gas of dense loops with weights $p_S$ or $1-p_S$
($S=A$, $B$) for every point where loops come together, and a
global weight factor ${\rm tr}\, (-1)^F=1$ for contractible loops.
The weight of non-contractible loops (necessarily in the time
direction, since the system has free boundary conditions in the
space direction) is $2n+1$. It would be $n+1-n=1$ if the partition
function were defined as the supertrace $\rm STr$ (or ${\rm
Tr}_m$), leading to the partition function $Z=1$. Note that for
free boundary conditions, a loop can wind the time direction at
most once.

The modified partition function with weight $2n+1$ per
non-contractible loop can be computed using a mapping onto the
6-vertex model. To do so, we observe that the loop model partition
function is the same as the well-known reformulation of the $Q$
state Potts model (with $Q=1$) partition function using the
polygon decompositions of the surrounding (or medial) lattice
\cite{bax}. The mapping onto the 6-vertex model then follows from
orienting the loops, and gluing them at the interaction vertices.
The resulting 6-vertex model has anisotropy parameter
$\Delta={1\over 2}$. The loop orientation can be thought of as an
isospin $S^z=\pm 1/2$, and the insertion of a term $k^{2S^z}$ in
the trace finally produces the right weight for non-contractible
loops provided
\begin{equation}
(2)_k =2n+1,
\end{equation}
where as usual we have set $(x)_k=(k^x-k^{-x})/(k-k^{-1})$. In the
following we also use the parameter $\alpha=\ln k$, such that
$(2)_k=2\cosh\alpha$.

The 6-vertex model has a Hilbert space ${\cal H}_{\rm 6-vertex}$
which is entirely unrelated to the one of the supersymmetric
model; in particular its dimension is  $2^{2l_1}$! The point
however is that the 6-vertex model is solvable, so all of the
transfer matrix eigenvalues can be obtained: these eigenvalues
will also be the ones of the initial supersymmetric problem, up to
multiplicities which we now elucidate.

To proceed, we recall that the 6-vertex model transfer matrix
enjoys a crucial property: it has a quantum group sl$(2)_t$
isospin symmetry with $t=e^{i\pi/3}$, $[T_{\rm 6-vertex},\,{\rm
sl}(2)_t]=0$. Since the theory of representations of sl$(2)_t$ for
$t$ a root of unity is a bit intricate, let us for a while assume
the vertex model is at some generic coupling, with $t$ not a root
of unity. In that case, the quantum group symmetry results in a
splitting of ${\cal H}_{\rm 6-vertex}$ into irreducible
representations of sl$(2)_t$ according to
\begin{equation}
{\cal H}_{\rm 6-vertex}=\sum_{j=0}^{l_1}
\left[\left(\begin{array}{c}2l_1\\ l_1-j\end{array}\right)-
\left(\begin{array}{c}2l_1\\ l_1-j-1\end{array}\right)\right]
\rho_{j},
\end{equation}
where $j$ is the sl$(2)_t$ isospin, the $\rho_j$ are sl$(2)_t$
representations, and the coefficients are binomial coefficients.
Then there are
\begin{equation}
\left(\begin{array}{c}2l_1\\ l_1-j\end{array}\right)-
\left(\begin{array}{c}2l_1\\ l_1-j-1\end{array}\right)
\label{numbin}
\end{equation}
representations of isospin $j$, each coming with its own
eigenvalue. The number (\ref{numbin}) is nothing but the dimension
of the irreducible representation $\Pi_j$ of the commutant algebra
of sl$(2)_t$, the Temperley-Lieb algebra.

The generating function of the eigenvalues of the transfer matrix
in this representation $Z_j=\sum_{\Pi_j}\lambda^{l_2}$  (that is,
the generating function of the eigenvalues at isospin $j$) can be
obtained by taking the generating function in the sector $S^z=j$
and subtracting the generating function in the sector of isospin
$S^z=j+1$: $Z_j=Z_{S^z=j}- Z_{S^z=j+1}$. Commutation with the
sl$(2)_t$ algebra ensures coincidence of the eigenvalues that have
to be subtracted, even in finite size.

The generating function of the 6-vertex model eigenvalues reads
then
\begin{equation}
Z_{\rm 6-vertex}=\sum_{j=0}^{l_1} (2j+1)\,Z_j.\label{zvertex}
\end{equation}
To obtain what will be the generating function of the sl$(n+1|n)$
model eigenvalues after the 6-vertex model parameters are adjusted
so $t=e^{i\pi/3}$, we need to give a weight $2n+1$ per
non-contractible loop, that is introduce the term $k^{2S^z}$ in
the trace.  Since the 6-vertex transfer matrix commutes with
sl$(2)_t$, this simply gives a multiplicity factor which is the
$k$-deformation of the one in (\ref{zvertex}):
\begin{equation}
Z_{\rm mod}={\rm Tr}\,T^{l_2}=\sum_{j=0}^{l_1} (2j+1)_k\,Z_j. %\label{susy}
\end{equation}
The decomposition of the sl$(n+1|n)$ Hilbert space in terms of
irreducible representations of the Temperley-Lieb algebra also
follows:
\begin{equation}
{\cal H}_{\rm susy}= \sum_{j=0}^{l_1} (2j+1)_k\,  \Pi_j.
\end{equation}
[For $n=1$, the first few multiplicities are $(1)_k=1$, $(3)_k=8$,
$(5)_k=55$, \ldots, and all are Fibonacci numbers.] The total
cardinality reads then
\begin{equation}
\hbox{Dim }{\cal H}_{\rm susy}=\sum_{j=0}^{l_1} (2j+1)_k
\left[\left(\begin{array}{c}2l_1\\ l_1-j\end{array}\right)-
\left(\begin{array}{c}2l_1\\ l_1-j-1\end{array}\right)\right].
\end{equation}
For future purposes, we note that the cardinality can also be
written
\begin{eqnarray}
\hbox{Dim }{\cal H}_{\rm susy}&=&\sum_{j=1}^{l_1}
\left(\begin{array}{c}2l_1\\ l_1-j\end{array}\right)
\left[(2j+1)_k-(2j-1)_k\right]\nonumber\\
&&\mbox{}+\left(\begin{array}{c}2l_1\\
l_1\end{array}\right)\nonumber\\ &=&2\sum_{j=1}^{l_1}
\left(\begin{array}{c}2l_1\\ l_1-j\end{array}\right)
\cosh2j\alpha+\left(\begin{array}{c}2l_1\\
l_1\end{array}\right)\nonumber\\ &=&
(e^{\alpha}+e^{-\alpha})^{2l_1}=(2n+1)^{2l_1}. \label{counting}
\end{eqnarray}
%
%The last line is elementary.
%For the value of $k$ relevant here,
%since $(2)_k=2n+1$, the dimension is $(2n+1)^{2l_1}$.

We now let $t=e^{i\pi/3}$. None of the previous results change,
except that ${\cal H}_{\rm 6-vertex}$ splits into indecomposable
representations of sl$(2)_t$ instead of irreducible ones.
Similarly, the representations of the Temperley-Lieb algebra have
to be handled more carefully. However, the arguments using the
counting of states still hold by continuity, and so do those that
lead to the partition functions in the continuum limit.

The conformally-invariant limit for the sl$(n+1/n)$ model is
obtained in the isotropic case when $p_S=1/2$. Up to a
non-universal bulk free energy term (which is trivial here,
because the unmodified partition function is $Z=1$), the
expression of $Z_j$ in the conformal limit follows from the
computations in \cite{sb}. In the limit where $t\rightarrow
e^{i\pi/3}$, one finds
\begin{equation}
Z_j\rightarrow {q^{j(2j-1)/3}-q^{(j+1)(2j+3)/3}\over P(q)}.
\end{equation}
Here, $P(q)=\prod_{n=1}^\infty (1-q^n)$, and $q=e^{-\pi l_2/l_1}$.
The generating function for the spectrum of the sl$(n+1|n)$ model
with free boundary conditions finally follows:
\begin{equation}
Z_{\rm mod}=\sum_{j=0}^{\infty} (2j+1)_k\,
{q^{j(2j-1)/3}-q^{(j+1)(2j+3)/3}\over P(q)}.
\end{equation}

The unmodified partition function is reproduced meanwhile by
taking
\begin{equation}
Z=\sum_{j=0}^{l_1} (2j+1)_t Z_j.
\end{equation}
as $t\rightarrow e^{i\pi/3}$. That it is equal to $1$ in that
limit is not obvious from this formula alone: to prove it, one
needs to invoke the additional cancellations that arise from
representation theory of sl$(2)_t$ when $t=e^{i\pi/3}$. These
cancellations completely mimic the ones of the continuum limit,
which give here
\begin{equation}
Z={1\over P(q)}\left(1-q-q^2+q^5+q^7-q^{12}+\ldots\right),
\end{equation}
which is equal to one by Euler's pentagonal number theorem.

Of course, the fact that $Z=1$ follows much more directly from the
fact that it coincides with the partition function of the
supersymmetric model with periodic boundary conditions in the time
direction.

In conclusion, we see that the eigenvalues of the transfer matrix
based on sl$(n+1|n)$ are the {\em same} as the eigenvalues of the
6-vertex transfer matrix. The only things that differ are
multiplicities, none of which are zero, i.e.\ any eigenvalue of
one transfer matrix is also an eigenvalue of the other.

The operator content of the sl$(n+1|n)$ model with free boundary
conditions is therefore made only of $j$ hull operators with
surface dimensions
\begin{equation}
h_j={j(2j-1)\over 3}
\end{equation}
and multiplicities (neglecting the effect of descendants this
time)
\begin{equation}
D_j=(2j+1)_k.
\end{equation}
These multiplicities are much larger than dimensions of
irreducible representations of sl$(n+1|n)$; the commutant of the
Temperley-Lieb algebra in the supersymmetric representation used
here is much larger than (the products of generators of) the
global supersymmetry sl$(n+1|n)$ \cite{rsunpub}. The values of
$D_j$ can however be understood by a simple argument: calling
$D_1=D_{\rm adj}=(2n+1)^2-1$ the dimension of the adjoint, one
finds that
$$ D_{j}=D_{\rm adj} \, D_{j-1}-D_{j-1}-D_{j-2}. $$
This is easily interpreted graphically: with a pair of non
contracted lines we associate the adjoint. Multiplying $2(j-1)$
non-contracted lines by the adjoint gives either $2j$ non
contracted lines, or $2(j-1)$ if one of the added lines gets
contracted with the ones existing before, or $2(j-2)$ if the two
added lines get contracted.

Here, it may be useful to comment briefly on the decomposition of
products of the type $(V\otimes V^\ast)^{l_1}$ in the case $n=1$.
Following notations of \cite{sorba}, one has for instance
\begin{eqnarray}
V\otimes V^\ast&=&\Pi(0,0)\oplus\Pi(0,1),\nonumber\\ (V\otimes
V^\ast )^2&=&\Pi(0,0)\oplus\Pi(0,2)\oplus
4\Pi(0,1)\oplus\Pi(1/2,3/2)\nonumber\\&&\mbox{}\oplus
\Pi(-1/2,3/2) \oplus[Indecomp.],
\end{eqnarray}
where in particular $\Pi(0,1)$ is the adjoint, and $[Indecomp.]$
stands for an eight-dimensional indecomposable block. Recall that
the representations $\Pi(b,j)$ are generically typical, with
dimension $8j$ and vanishing supertrace, while $\Pi(\pm j,j)$ are
atypical, with dimension $4j+1$, and supertrace equal to $\pm 1$.
We observe in the previous examples that the ``true'' singlet
appears only once: it can be proved that this is a general feature
(invariant vectors, that are annihilated by all the generators,
appear in the indecomposable blocks in general, but they result
from the action of some generators on other vectors inside the
block). This is compatible (though not necessarily equivalent)
with having $Z=1$. With our normalizations for the transfer
matrix, the eigenvalue associated with the true singlet is
$\lambda=1$.

A similar approach could presumably be used to study periodic
boundary conditions. Commutation with the quantum group sl$(2)_t$
would then be replaced by the existence of commutative diagrams as
discussed in \cite{ps}. We are not going to discuss this in detail
here, but would like to indicate how the counting of states works
in that case.

The modified partition function of the sl($n+1|n$) model with
periodic boundary condition in the space direction, $Z_{\rm mod}$,
will be obtained by combining partition functions of the 6-vertex
model with a given value of the isospin $S^z=M$, and a twist angle
$\phi$ (we use the notations of \cite{ps}). This corresponds in
the Coulomb gas at coupling $gf^2=g/4=2/3$, to $m=S^Z=M$,
$e=\phi/2\pi +P$, where $P$ is an arbitrary integer. The 6-vertex
partition function in the continuum limit is
\begin{equation}
Z_{S^Z}^{\phi/2\pi}={1\over \eta(q)\eta(\overline{q})}\sum_P
q^{\Delta_{\phi/2\pi+P,S^z}(g/4)}
\overline{q}^{\overline{\Delta}_{\phi/2\pi+P,S^z}(g/4)}.
\end{equation}
The partition function of the supersymmetric model for a given
value of $M$ reads meanwhile, for $M\neq 0$,
\begin{eqnarray}
Z_{\pm M}
&=&\sum_{d'}G(M,d';1/2,e_0,2\alpha)\frac{1}{\eta(q)\eta(\overline{q})}\nonumber\\
&&\mbox{}\times\sum_{P'}
q^{\Delta_{P'/d',M}(g/4)}\overline{q}^{\overline{\Delta}_{P'/d',M}(g/4)},
\end{eqnarray}
The sum over $P'$ can be decomposed into a sum of $d'$ terms for
all the congruences modulo $d'$; it follows that
\begin{equation}
Z_{\pm M}=\sum_{d'} G(M,d';1/2,e_0,2\alpha)\sum_{k=0}^{d'-1}
Z_M^{k/d'}.
\end{equation}
It is very likely that the same expression holds in the finite
case as well. If we assume this is true, we can check that the
right number of states for the supersymmetric model is obtained.
Indeed, each $Z_{S^z}^{\phi/2\pi}$ being a partition function of
the 6-vertex model at isospin $S^z$ accounts for
$\left(\begin{array}{c} 2l_1\\ l_1-S^z\end{array}\right)$ states,
independently of the value of the twist angle $\phi$. Hence the
number of states contributing to $Z_{\pm M}$ is
$$
 \left(\begin{array}{c} 2l_1\\
l_1-M\end{array}\right)\sum_{d'} d'~G(M,d';1/2,e_0,2\alpha). $$
Now,
\begin{eqnarray}
\lefteqn{\sum_{d'}d'~G(M,d';1/2,e_0,2\alpha)}
&&\nonumber\\&\quad=&2\sum_{d',d} w(M,d;1/2,e_0,2\alpha)
\mu(d'/d).
\end{eqnarray}
Recall that $d|M,d'|M$. Then $\sum_{d'}\mu(d'/d)=\delta_{d,M}$
from Eq.\ (\ref{mobius1}), and the right hand side reduces to
$2\cosh 2M\alpha$. It follows that the number of states
contributing to $Z_{\pm M}$ is $2\cosh 2M\alpha
\left(\begin{array}{c} 2l_1\\ l_1-M\end{array}\right)$, for $M\neq
0$. The case $M=0$ is trivial: the corresponding partition function of
the supersymmetric model has the form (\ref{mequalzero}), and is
reproduced in the 6-vertex model by taking $S^z=0$ and
$\phi/2\pi=e_0$, with a corresponding number of states equal to
$\left(\begin{array}{cc}
2l_1\\
l_1\end{array}\right)$. For a
system of size $2l_1$, the allowed values of the isospin run
between $0$ and $l_1$, so using (\ref{counting}) we find the right
size for $\hbox{Dim }{\cal H}_{\rm susy}$.

Notice that in finite size, there will be numerous coincidences of
levels ensured by the commutative diagrams of \cite{ps}. These
should guarantee that subtractions of partition functions are
possible in the finite size case as well, like in the case of free
boundary conditions.

We should point out that the arguments given in this Appendix
generalize to other $m$ and $n$ values, including the range $m>2$
where the models are gapped, not conformal. This includes the
cases with $n=0$, the SU($m$) spin chains with alternating $n$ (or
$V$) and $\overline{n}$ ($V^\ast$) representations. For
$\epsilon=1$, the eigenvalues of these models can be found from
the Bethe ansatz on the corresponding 6-vertex model \cite{ps},
and the multiplicities can now be found as here.

\end{document}